\documentclass[aps,twocolumn,prl,amsfonts,showpacs,superscriptaddress,longbibliography]{revtex4-2}
\DeclareUnicodeCharacter{0393}{$\Gamma$}
\usepackage{verbatim,epsfig,amsmath,amssymb,bm,epsf,graphicx,psfrag,bbold,amsthm,amsfonts}
\usepackage[bottom]{footmisc}
\usepackage{hyperref}
\usepackage{cleveref} 
\usepackage{enumitem}
\usepackage{framed}
\usepackage{mathrsfs}
\usepackage{esint}
\setlist{nosep}
\usepackage{placeins}
\usepackage[all]{xy}
\usepackage{color}
\usepackage[utf8]{inputenc}
\usepackage{float}
\usepackage{natbib}
\usepackage{tikz-cd}
\usepackage{verbatim}
\usepackage{leftidx}
\usepackage{physics}
\usepackage{ulem}

\usepackage{pifont}
\usepackage{braket}
\usepackage{booktabs}

\newcommand{\Gb}{\mathbf{G}}
\newcommand{\rb}{{\bm r}}
\newcommand{\kb}{{\bm k}}

\newcommand{\aB}{a_{\scriptscriptstyle B}}

\begin{document}


\title{Electron quasicrystal in semiconductor quantum wells}

\author{Filippo Gaggioli}
\affiliation{Department of Physics, Massachusetts Institute of Technology, Cambridge, MA-02139, USA}

\author{Pierre-Antoine Graham}
\affiliation{Department of Physics, Massachusetts Institute of Technology, Cambridge, MA-02139, USA}

\author{Liang Fu}
\affiliation{Department of Physics, Massachusetts Institute of Technology, Cambridge, MA-02139, USA}

\date{\today}

\begin{abstract}
The homogeneous electron gas is a cornerstone of quantum condensed matter physics, providing the foundation for developing density functional theory and understanding electronic phases in semiconductors. 
However, theoretical understanding of strongly-correlated electrons in realistic semiconductor systems remains limited. In this work, we develop a neural network based variational approach to study quantum wells in three dimensional geometry for a variety of electron densities and well thicknesses. Starting from first principles and without pre-assumed electron ordering, our  method reveals metallic and crystalline phases with both monolayer and bilayer charge distributions.
In the emergent bilayer, we discover a new quantum phase of matter: the electron quasicrystal.
\end{abstract}
\maketitle





The success of condensed matter physics rests on the availability of archetypal models that distill the essential physics of diverse electronic materials. Among these, the homogeneous electron gas holds a central place. Theory of Coulomb gas provides the conceptual basis of density functional theory \cite{Kohn_1964, Kohn_1965, Perdew_2003} and underpins our understanding of low-density electrons in semiconductors.
Its two-dimensional realization in semiconductor quantum wells and atomically thin materials  also provides the platform for the quantum Hall effect, where Coulomb interactions and Landau quantization produce incompressible quantum liquids \cite{Laughlin_1983}.

Recently, a robust fractional quantum Hall state with energy gaps up to $6$K has been observed \cite{Singh_2024,Singh_2025} at filling factor $\nu=1/2$ in ultra-clean GaAs quantum well \cite{Chung_2021}. This state---a potential host of non-Abelian anyons \cite{Moore_1991, Nayak_2008, Sharma_2024}---contrasts with the composite Fermi liquid state commonly seen at $\nu=1/2$, and it appears only in {\it wide} quantum wells.
The increased well thickness introduces a new energy scale: the subband spacing between quantized transverse modes. 
The interplay of kinetic energy, Coulomb repulsion, and the subband energy gap enriches the physics of two-dimensional electron gas (2DEG) \cite{Drummond_2009, Smith_2024} and gives rise to quantum phases with no analog at zero layer thickness \cite{MacDonald_1990, Boebinger_1990, Eisenstein_1992, Narasimhan_1995, Manoharan_1996,Eisenstein_1992, Hatke_2015}. Notably, the 2DEG in a {\it single} wide quantum well can develop a bilayer-like charge distribution while retaining significant interlayer tunneling \cite{Suen_1991}. Intriguingly, the $\nu=1/2$ FQH state only emerges near the transition from a monolayer to a bilayer regime \cite{Singh_2024,Singh_2025}.   

Theoretical understanding of quantum phases of 2DEG in realistic quantum wells with finite thickness remains limited. Conventional approaches such as density functional theory or Hartree-Fock theory fail to capture strong correlation effect at low electron density, while exact diagonalization and density-matrix renormalization group methods have difficulty in treating continuum systems in three-dimensional geometry. Recent advances in machine learning \cite{Carrasquilla_2017, Carleo_2017} have opened a new route for solving the ground states of strongly interacting electronic systems in continuous space 
\cite{ Cassella_2023, Li_2022, Wilson_2023, Pescia_2024, Geier_2025}.
In these approaches, many-body wavefunctions are represented by deep neural networks whose large number of variational parameters can be efficiently optimized via energy minimization within the variational Monte Carlo framework.
This neural-network variational Monte Carlo (NN-VMC) method has already demonstrated remarkable accuracy across a diverse set of strongly correlated systems, including metal–insulator transitions in twisted transition-metal dichalcogenides \cite{Geier_2025}, fractional quantum Hall states \cite{Teng_2025, Nazaryan_2025}, and chiral superconductivity \cite{Li_2025}.


In this work, we employ unbiased attention-based NN-VMC to quantitatively determine the many-body ground states of interacting electrons in a semiconductor quantum well from first principles.
Our calculations reveal a rich phase diagram---parametrized by the electron density and well thickness---that includes metallic phases as well as electron crystals in both monolayer and bilayer regimes.  
We demonstrate that increasing the well thickness at fixed electron density enhances correlation effects, thereby promoting crystal states and opening a promising route toward their experimental realization.

Interestingly, in the bilayer regime, our NN-VMC uncovers a previously unknown electronic state of matter, which we term the \textit{quantum electron quasicrystal}. This state exhibits quasicrystalline charge order \cite{Shechtman_1984, Levine_1984, Goldman_1993_review} and has no classical analogue: its stability relies on quantum fluctuations.
We support this conclusion by analyzing the energetics of bilayer electron crystals and quasicrystals and highlighting the importance of zero-point quantum fluctuations.
The quantum electron quasicrystal thus represents one of the first clear examples of a  new quantum phase of matter discovered through neural-network--powered many-body calculation. Our theoretical predictions can be readily tested in semiconductor quantum wells and van der Waals heterostructures \cite{Smolenski_2021, Zhou_2021, Li_2021,Li_2024}.

\begin{figure}
    \includegraphics[width=0.95\linewidth]
    {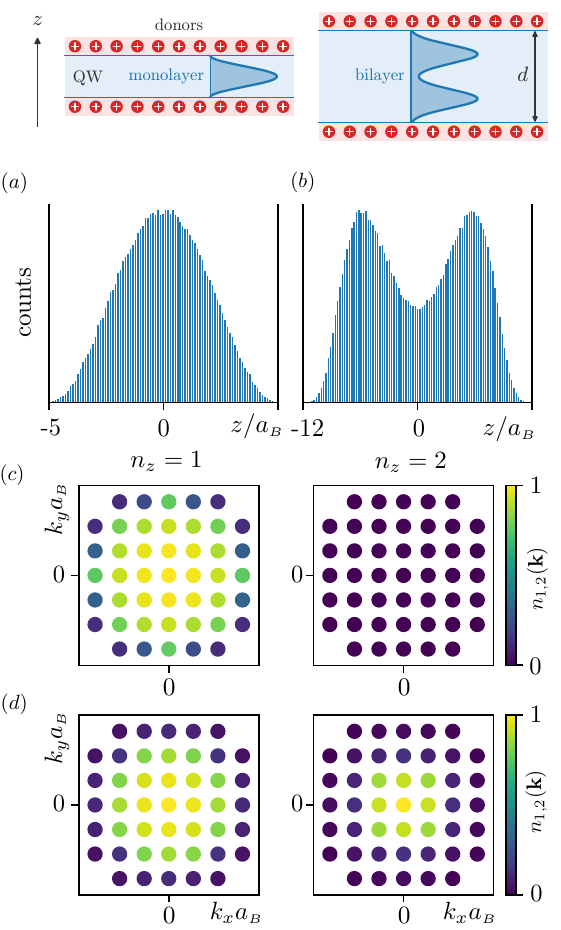}
    \caption{\textbf{Monolayer and bilayer metals:}  
    Top: quantum well schematics, with donor atoms forming a positive-charge background at the top and bottom interfaces. Out-of-plane charge density distribution
    at $(r_{2D}/\aB, d/\aB) = (10, 10)$ $(a)$ and $(10, 25)$ $(b)$.
    In the mononolayer metal all the electrons live in the lowest subband $n_z = 1$, giving rise to a single Fermi surface in the $2$D momentum space occupation $n_{1}(\mathbf{k})$ obtained from the $1$-RDM $(c)$.
    In the bilayer metal, electrons  are distributed between multiple subbands (for this specific example, $n_z=1,2$) and their momentum-space occupations $n_{1}(\mathbf{k})$ and 
    $n_{2}(\mathbf{k})$ form distinct Fermi surfaces $(d)$ Plots $(a)-(d)$ were obtained from NN-VMC simulations of Eq.~\eqref{eq:Hamiltonian} for $N=30$ electrons using a square periodic supercell.}
\label{fig:mono_to_bilayer}
\end{figure}

\textit{Model --} Our system consists of spin-polarized electrons, as realized under an in-plane magnetic field, confined in a quantum well with thickness $d$ along the out-of-plane direction $z$.
In effective atomic units, this is described by the dimensionless model Hamiltonian: 
\begin{equation}\label{eq:Hamiltonian}
    H = \sum_i\left( -\frac{1}{2}\nabla_i^2 + V(z_i) \right) + \frac{1}{\aB}\sum_{j<i}\frac{1}{|\rb_i-\rb_j|} ,
\end{equation}
after introducing the effective Bohr radius $\aB= 4\pi\epsilon\hbar^2/m e^2$, with $e$ the electron charge, $m$ the effective mass, and $\epsilon$ the dielectric constant. 
The confining potential $V(z)$ provides a realistic description of semiconductor quantum wells \cite{Singh_2025} and consists of  two contributions: 1. a potential step at $z=\pm d/2$ corresponding to the difference in the conduction band energies of the well and the barrier, e.g., GaAs and AlGaAs; 2. an electrostatic potential created by two layers of neutralizing positive charges located outside the well (cf schematics at the top of Fig.~\ref{fig:mono_to_bilayer}). $V(z)$ is  flat inside the quantum well, rises steeply at the boundaries $z \approx \pm d/2$ and increase linearly outside, see the Supplementary Material (SM) \cite{supplementary} for more details.

\textit{Neural network VMC --} 
Determining the ground state of the interacting Hamiltonian \eqref{eq:Hamiltonian} is a formidable task because of the high-dimensional continuous configuration space. We tackle this by generating many-electron variational wavefunctions $\psi_{\scriptscriptstyle \theta}(\rb_1,\ldots,\rb_N)$ from a trainable, attention-based  neural network  \cite{vonGlehn_2023}, with its weights and biases $\theta$ being variational parameters \cite{Foster_2025, Geier_2025, Gaggioli_2026}. To implement the slab geometry, we adopt a feature layer that enforces periodic boundary condition for electron's in-plane coordinates, while leaving the out-of-plane coordinate along $z$ open
, see the SM \cite{supplementary} for details.

Having constructed the variational wave function, we train $\psi_{\scriptscriptstyle \theta}$ by minimizing the energy of the system via a VMC approach: the expectation value of the Hamiltonian \eqref{eq:Hamiltonian} is evaluated by means of Monte Carlo averaging and used to generate the gradients $\nabla_\theta H$ that are passed back to the KFAC optimizer \cite{martens2015optimizing} to update the weights $\theta$ of the neural network via standard backpropagation after each training step.

\textit{Results --} We now present the results of our NN-VMC simulation and analyze the phase diagram of Hamiltonian \eqref{eq:Hamiltonian} as a function of the electron density and the quantum well thickness.

In the two-dimensional limit $d\to 0$, electrons are tightly confined to the $z=0$ plane and the resulting 2D Hamiltonian \eqref{eq:Hamiltonian} displays only two energy scales, $E_k \sim r_{2D}^{-2}$ and $E_c \sim r_{2D}^{-1}$, with $r_{2D} = (\pi\,n_{2D})^{-1/2}$ and $n_{2D}$ the areal density. As is well known, such 2DEG is then characterized by a single dimensionless parameter $r_s \equiv r_{2D}/a_B$; the ground state is a Fermi liquid at high density and a Wigner crystal at low density, with the transition occurring at $r_s \sim 30$.

The well thickness introduces a third energy scale $E_w \sim d^{-2}$, corresponding to the energy gap separating the different \textit{subbands} associated with the quantized motion along $z$. In narrow wells $E_w$ is the largest energy scale and only the lowest subband (quantum number $n_z = 1$) is occupied. As a result, the charge density profile is monolayer, 
as shown in Fig.\ \ref{fig:mono_to_bilayer}$\,(a)$ for our NN-VMC simulations. 
Correspondingly, the subband-resolved momentum space occupation $n_{1}(\mathbf{k})$ extracted from the one-body reduced density matrix (1RDM) (cf SM \cite{supplementary})) displays a single, sharp Fermi surface, characteristic of the metallic state (see Fig.\ \ref{fig:mono_to_bilayer}$\,(c)$).

In wider quantum wells, when $E_w$ becomes comparable to the kinetic and Coulomb energies, higher subbands become populated, which allows electrons to spread out along $z$ and stay away from each other. 
Indeed, the charge density distribution develops into the bilayer profile displayed in Fig.\ \ref{fig:mono_to_bilayer}$\,(b)$ and, at sufficiently large densities, both the lowest and first excited subbands ($n_z=1,2$) are occupied. The presence of two Fermi surfaces of different sizes in $n_{1}(\mathbf{k})$ and $n_{2}(\mathbf{k})$, shown in Fig.\
\ref{fig:mono_to_bilayer}$\,(d)$, leads to two different frequencies in quantum oscillation as observed experimentally \cite{Jungwirth_1997}. 
We note that, while the emergence of a bilayer charge distribution in a single quantum well can be qualitatively described by density functional theory or mean field approaches 
\cite{Abolfath_1997, Sharma_2024, Singh_2025}, these methods fail to capture strong correlation effects at small densities, as we will show below.

\begin{figure}
    \includegraphics[width=0.9\linewidth]    {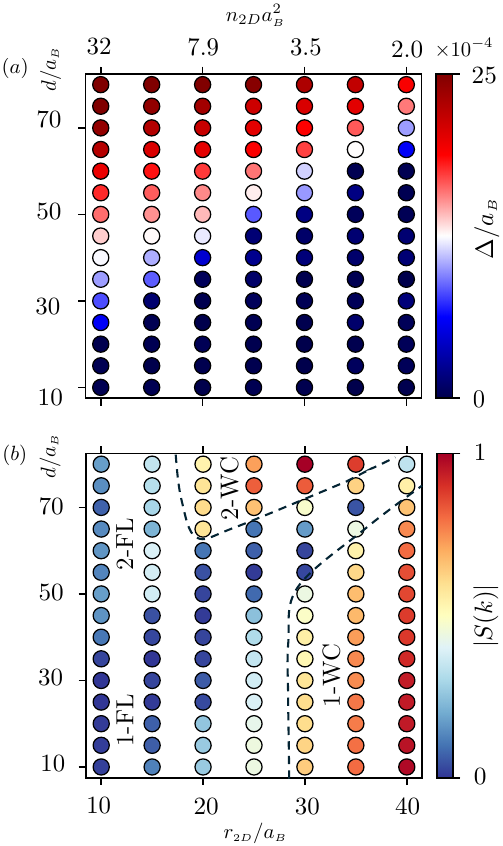}
    \caption{\textbf{Phase diagram: }  
    $(a)$ Out-of-plane charge distribution as a function of the electron density and the quantum well thickness. Blue and red dots correspond to monolayer (one peak at $z= 0$, e.g., Fig.\ \ref{fig:mono_to_bilayer}$\,(a)$) and bilayer states (two peaks spaced by a finite $\Delta$, e.g., Fig.\ \ref{fig:mono_to_bilayer}$\,(b)$), respectively.
    $(b)$ In-plane charge distribution,
    as characterized by the average height of the structure factor   $S(\kb)$ (Bragg) peaks.
    In the Fermi liquid phase (cool colors), the charge distribution is homogeneous and the structure factor does not have sharp peaks. In the crystalline phase (warm colors), the charge arrangement gives rise to strong peaks in $S(\kb)$ (cf insets in Fig.\ \ref{fig:bilayer_WC}).
    Dashed lines qualitatively separate the mono- ($1$-WC) and bilayer ($2$-WC) crystals from the Fermi liquid regions ($1$-FL and $2$-FL). The phase diagrams $(a)$ and $(b)$ were obtained for $N=30$ particles and square supercell geometry.}
    \label{fig:phase_diagrams_combined}
\end{figure}

The phase diagram in Fig.\ \ref{fig:phase_diagrams_combined}$\,(a)$ provides an overview of the charge density profile as a function of electron density and the well thickness, with the blue and red dots representing monolayer and bilayer charge distribution.
The bilayer spacing $\Delta$ grows monotonically upon increasing the electron density or the well thickness. Such bilayer charge distribution is induced by electron-electron interaction in a wide quantum well; it does not occur in the noninteracting limit even when higher subbands are occupied. 

Notably, increasing quantum well thickness has multifaceted effects on the competing ground states of interacting electrons. 
Compared with the strictly two-dimensional limit ($d=0$), a small well thickness leads to the smearing of the electron wave-packet in the out-of-plane direction, which effectively softens the Coulomb energy to $E_c\sim (r_{2D}^2 + d^2)^{-1/2}/\aB$ \cite{Zhang_1986}. The critical value of $r_{2D}/\aB$ for Wigner crystallization therefore increases with thickness $\propto(1 + d^2/r_{2D}^2)^{-1/2}$.

On the other hand, in the bilayer regime where the total charge splits equally on two sides of the well, the in-plane interparticle distance within each layer is reduced by $\sim 1/\sqrt{2}$, leading to enhanced ratio of interaction to kinetic energy. This allows for a bilayer crystal to form already at electron areal densities well above the critical density $r_s\sim 30$ for spin-polarized Wigner crystallization in strictly two dimensions.  

An overview of the crystalline phase diagram for $N=30$ particles (square supercell) is shown in Fig.\ \ref{fig:phase_diagrams_combined}$\,(b)$, where warmer colors indicate stronger Bragg peaks in the amplitude of the in-plane structure factor $S(\kb) \propto \langle \sum_{i,j}e^{-i\kb\cdot(\rb_i - \rb_j)} \rangle$. The monolayer electron crystal is realized at low densities $r_s \gtrsim 30$, while the bilayer crystal is present already at larger densities $r_s \gtrsim 20$ for well thicknesses $d/\aB \gtrsim 60$. 
Remarkably, the two crystalline phases are separated by an intervening Fermi Liquid phase (blue dots) that follows the monolayer-bilayer transition line in Fig.\ \ref{fig:phase_diagrams_combined}$\,(a)$. Thus, in a sufficiently thick well, the system undergoes a sequence of phase transitions upon reducing the density: bilayer Fermi liquid $\rightarrow$ bilayer crystal $\rightarrow$ monolayer Fermi liquid $\rightarrow$ monolayer Wigner crystal. In particular, the bilayer crystal that intervenes between two metallic phases should be manifested as a resistance peak at {\it intermediate} densities. 

%

%
%

%
\begin{figure*}
    \includegraphics[width=0.99\linewidth]
    {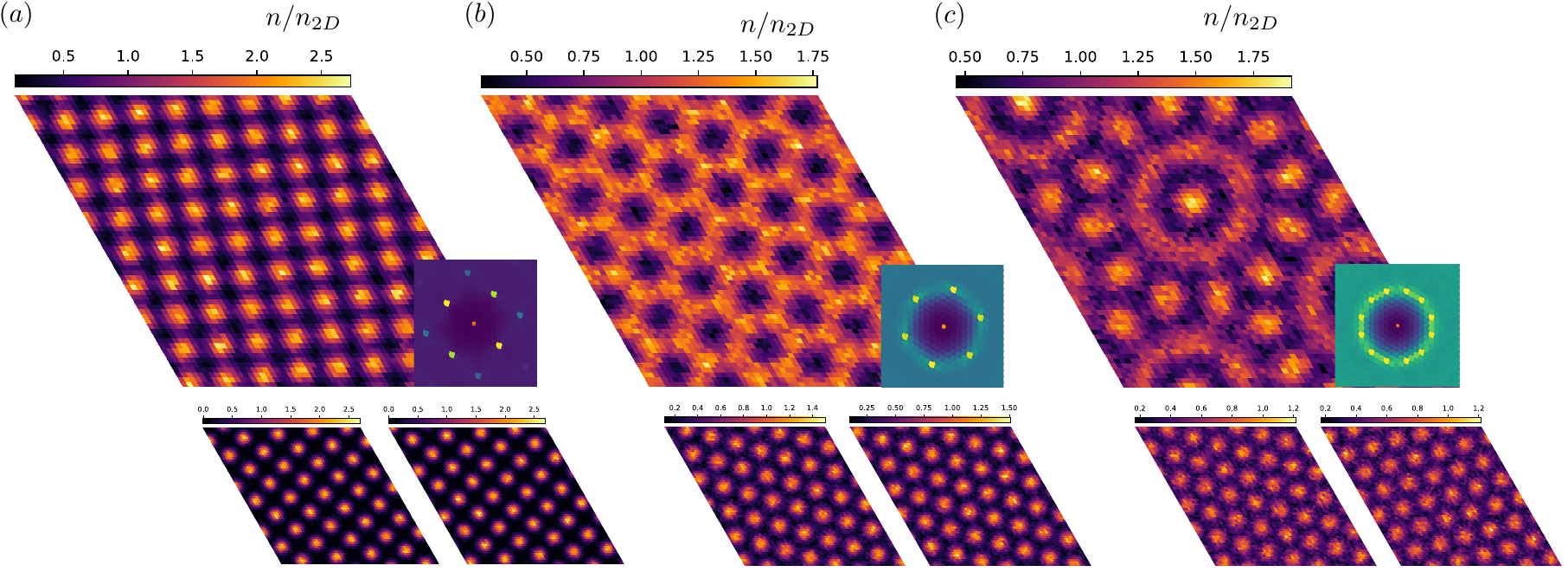}
    \caption{\textbf{Bilayer (quasi-)crystal:}  
    Square $(a)$, honeycomb $(b)$ and quasicrystal $(c)$ bilayer, corresponding to the ground state for $N_e = 78$ electrons (triangular supercell) at $(r_{2D}/\aB, d/\aB) = (40, 80), (30, 90)\text{ and } (30, 150)$, respectively.
    The commensurate stackings $(a)$ and $(b)$ can be understood from minimizing the intra-layer and inter-layer Coulomb repulsion. The electronic quasicrystal, on the other hand, does not have a classical analogue and is stabilized purely by quantum fluctuations, via the zero-point energy contribution of its extensive low-lying excitations.
    The bottom panels show the per-layer charge distribution in the spontaneously formed bilayer, while the colored insets display the Bragg peaks of the in-plane structure factor $S(\kb)$. The occurrence of twelve Bragg peaks in $(c)$ highlights the twelve-fold rotational symmetry of the quantum electron quasicrystal. 
    }
    \label{fig:bilayer_WC}
\end{figure*}

We now turn to the lattice structures of electronic crystals. In the monolayer regime, our NN-VMC finds a triangular lattice as in the 2D Wigner crystal at $d=0$. On the other hand, in the bilayer regime, a variety of charge orders appears, shown in Fig.\ \ref{fig:bilayer_WC} for $N= 78$ particles (triangular supercell) and different choices of parameters $(r_{2D}/\aB, d/\aB)$.
The square $(a)$ and honeycomb $(b)$ configurations are realized at small and intermediate well thicknesses respectively, which correspond to rotationally aligned stacking of square or triangular lattice crystals on the top and bottom layers, laterally shifted to minimize the Coulomb repulsion between the layers. These results also agree with Coulomb-induced electron crystals in the classical limit $\hbar \rightarrow 0$, see  Refs.~\cite{Goldoni_1996,WignerQuantumBilayer} and the
 SM \cite{supplementary}. 

For sufficiently large well thickness, on the other hand, our NN-VCM finds an unprecedented electronic state, characterized by a quasicrystalline charge order (Fig.\ \ref{fig:bilayer_WC}$\,(c)$) with the twelve-fold rotational symmetry highlighted by the structure factor $S(\kb)$ peaks (inset). This structure corresponds to $\approx30^\circ$ twisted stacking of two triangular lattice electron crystals. 





\begin{figure}
    \includegraphics[width=0.9\linewidth]{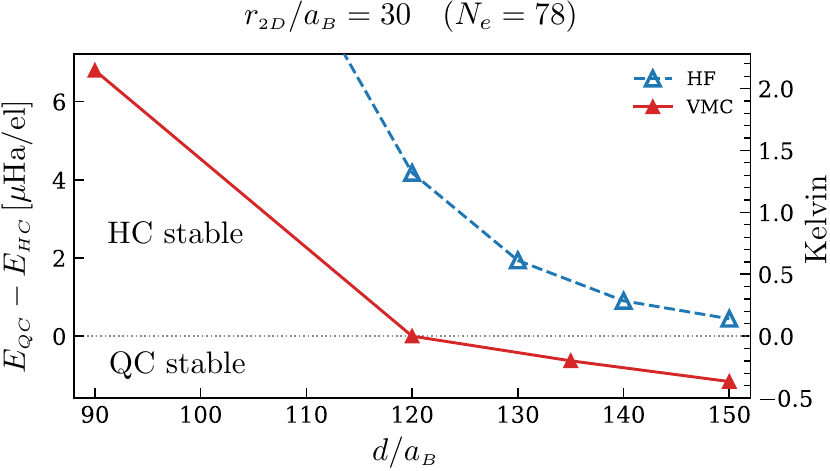}
    \caption{\textbf{Systematic energy comparison: }  
    NN-VMC energy difference (red line) between the honeycomb (HC) and quasicrystal (QC) stackings, for a triangular supercell with $N=78$ particles at fixed $r_{2D}/\aB=30$. The Hartree-Fock results for the same system  are shown with blue dashed line. Absolute energies are reported in the SM \cite{supplementary}.
}\label{fig:energy_comparison}
\end{figure}

We find this quasicrystalline ground state robustly across supercell geometries and system sizes. For example with periodic boundary conditions, the triangular supercell with $N=78$ electrons is commensurate with a twisted bilayer with angle $\theta=\arccos(23/26)\approx 28^\circ$ \cite{Pleasants_1996,Mele_2010} close to $30^\circ$. 
In Fig.~\ref{fig:energy_comparison}, we show the NN-VMC energy difference per electron between the honeycomb (HC) and quasicrystal (QC) stackings for a triangular supercell with $N=78$ electrons, as a function of $d/r_{2D}$ at fixed $r_{2D}/\aB=30$. This confirms that the quasicrystal becomes the true ground state for sufficiently large quantum well thickness, with a gap of the order of few hundred milliKelvin.
The absolute values of the energies in the HC and QC configurations, along with additional results for the case of a square supercell and for  twisted boundary conditions, are reported in the SM \cite{supplementary}.

Interestingly,  the quasicrystal state is not favored within Hartree-Fock calculations (HF, blue dashed curve in Fig.~\ref{fig:energy_comparison}), which neglect correlation effects: specifically, we find that HF energy lie significantly higher than NN-VMC energy throughout the phase diagram, and in this mean-field approximation the quasicrystal state is always higher in energy than the honeycomb (see SM \cite{supplementary}). 

As we show below, the \textit{electron quasicrystal} does not have a classical counterpart, as no incommensurate stacking can minimize the repulsive interaction between the layers.
Instead, its origin is quantum mechanical and can only be understood by accounting for quantum fluctuation in electrons' positions.

\textit{Quantum quasicrystal --} We first show that twisted bilayers with incommensurate structure---such as $30^\circ$-twist quasicrystal \cite{Ahn_2018, Yu_2019, Moon_2019, Uri_2023, Xia_2025}---cannot be the lowest energy stacking configuration in the classical limit $\hbar=0$. 
The classical energy is a sum of electrostatic interaction energies on the same layer and between the two layers. The intralayer energy is independent of the twist angle. The interlayer energy 
takes the general form $E_c =\sum_{l\neq l'} \int d\bm r  \int d \bm r' V(\bm r - \bm r' ) \rho_l (\bm r) \rho_{l'}(\bm r)$ with $\bm r$ the in-plane coordinate and $\rho_l$ the 2D charge density on layer $l$. When the electron crystal is formed,  $\rho_l = \rho_0 +\delta \rho_l$ becomes periodically modulated: $\delta \rho_l(\bm r) = \sum_{\Gb_l \neq 0} \rho_{\Gb_l} e^{i \Gb_l \cdot \bm r}$ with the reciprocal vectors $\Gb_l$. When two identical layers of electron crystals are stacked with a twist angle $\theta$, the interlayer energy $E_c$ vanishes in the thermodynamic limit if the bilayer structure is incommensurate ($\Gb_1 \neq \Gb_2$). Instead, at $\theta=0$ corresponding to a commensurate bilayer with  $\Gb_1=\Gb_2\equiv \Gb$, the interlayer interaction is generally nonzero and depends periodically on the lateral shift $\bm s$: $E_c(\bm s) = E_0 + \delta E_c(\bm s)$ with 
\begin{equation}\label{eq:interaction_energy}
\delta E_c(\bm s) = \sum_{ \Gb \neq 0} V_{\Gb} \rho_{\Gb} \rho_{-\Gb} e^{i \Gb \cdot \bm s}.  
\end{equation}
Since $\int d \bm s \; \delta E_c(\bm s)=0$, $\delta E_c$ must be negative at some optimum shift (except the singular case $\delta E_c=0$ for all shifts). Therefore, in the classical limit, the rotationally aligned bilayer is in general energetically preferred over incommensurate configurations (see SM \cite{supplementary} for details).

On the other hand, incommensuration has profound effects on the low-energy excitations that play an important role in shaping the quantum ground state of the bilayer, as we will now discuss. 
In a commensurate stacking, the two layers are locked together by Eq.\ \eqref{eq:interaction_energy} and, as a result, the ``out-of-phase" phonon modes acquire a finite gap:  
$\omega(k) = \sqrt{c^2k^2 + \omega_0^2}$, with $\omega_0^2 \sim \delta E_c/mr_{2D}^2$ at large $d$ \eqref{eq:interaction_energy} (we reinstated the electron mass $m$ for clarity).

In incommensurate stackings, by contrast, interlayer interactions average to zero over a lengthscale $\lambda\sim |\Gb_1 - \Gb_2|^{-1}$ set by the mismatch in the layers' reciprocal lattice vectors. On larger distances, the two layers are effectively decoupled yielding that out-of-phase phonons retain a linear dispersion $ck$ for wave vectors $k \lesssim 1/\lambda$, instead of acquiring an optical phonon gap.  
This momentum cutoff 
is the largest in the case of $30^\circ$-twist quasicrystal, where $|\Gb_1-\Gb_2|$ is maximal and $\lambda \sim r_{2D}$ becomes comparable to the interparticle distance, thus making the self-averaging effective at finite system sizes. see SM \cite{supplementary} for details.

Owing to its large number of low-lying excitations, the quasicrystal configuration may be stabilized by quantum fluctuations, because its zero-point energy is lower than that of a crystal by an amount
 $E_0 \sim \hbar\!\int\! d^2\kb\,(\omega(\kb) - c|\kb|)$. 
%
Beyond the semiclassical harmonic approximation, non-perturbative effects such as interlayer tunneling and cooperative ring exchange may also affect the energetics \cite{Kivelson_1986, Sagi_2016, Kim_2022}. While a complete microscopic theory of electron quasicrystal remains to be developed, our analysis above identifies the quantum origin of the $30^\circ$-twisted quasicrystal state found by our NN-VMC simulations. 

\textit{Discussion --} 
In this work, we have employed NN-VMC techniques to map out the phase diagram of the homogeneous electron gas in realistic quantum wells with finite thickness by solving the many-body Schrodinger equation from first principles. Starting with a randomly initialized network without any prior knowledge, our method uncovers a wealth of quantum phases, both metallic and crystalline, with monolayer and bilayer charge distributions. 

In the bilayer regime, we have reported unbiased evidence for a new quantum phase of matter -- the electronic quasicrystal.
Its dodecagonal pattern can be understood as a two-dimensional cut of a four-dimensional periodic structure \cite{Kalugin_1985}, and thus provides a material platform for studying quantum phenomena in higher-dimensional crystals \cite{Nuckolls_2025}, such as the $4$D quantum Hall effect \cite{Kraus_2013, Lohse_2018}.

Typical values of the Bohr radius in semiconductors range from a few to tens of nanometers, placing the density and thickness regimes explored in this work well within current experimental reach.
The predicted crystal and quasicrystal phases can be probed via transport measurement  \cite{Manoharan_1996, Yoon_1999} and optical detection of exciton umklapp scattering \cite{Smolenski_2021, Zhou_2021}.
Their insulating nature should be manifested by a resistivity peak at intermediate densities, separated from the low-density monolayer Wigner crystal by an intervening bilayer metallic phase (cf. Fig.\ \ref{fig:phase_diagrams_combined}). 
In TMD heterostructures, direct STM imaging \cite{Li_2021, Tsui_2024} provides a powerful tool for detecting square, honeycomb and $30^\circ$-twisted bilayer structures.

While our NN-VMC simulation targets the ground state, the electronic quasicrystal can be stabilized at finite temperatures by entropic contributions  to the free energy from the extensive low-lying excitations. 
Similar entropic mechanism has been invoked to explain quasicrystalline order in soft-matter systems \cite{Haji-Akbari_2009, Iacovella_2011, Fayen_2024}.


Our work demonstrates the success of neural-network--powered methods in studying real-world quantum materials and quantum devices, which opens the door to new research directions  in the field of semiconductors and beyond.
Of immediate interest are strongly correlated electron systems with nontrivial band geometry 
and the exotic $\nu=1/2$ FQH state \cite{Peterson_2010, Thiebaut_2015, Zhu_2016}
observed in wide quantum wells under strong magnetic fields.

Looking ahead, we believe that our accurate NN-VMC technique based on first principles will provide a powerful tool for AI-aided design in semiconductor and quantum industry. In particular, the creation of digital twins for quantum hardware will have the potential to accelerate the growth of nascent quantum computing technologies.

\textit{Acknowledgments --} It is a pleasure to thank Mansour Shayegan and Leonid Levitov for stimulating and insightful discussions.  This work was supported by the National Science Foundation (NSF) Convergence Accelerator Award No. 2235945. 
We acknowledge MIT SuperCloud, the Lincoln Laboratory Supercomputing Center and IAIFI (the NSF Cooperative Agreement PHY-2019786) for providing computing resources that have contributed to the research results reported within this paper.  
F.G. is grateful for the financial support from the Swiss National Science Foundation (Postdoc.Mobility Grant No. 222230).  
L.F. was supported by a Simons Investigator Award from the Simons Foundation.

\bibliography{biblio.bib}

\begin{thebibliography}{72}%
\makeatletter
\providecommand \@ifxundefined [1]{%
 \@ifx{#1\undefined}
}%
\providecommand \@ifnum [1]{%
 \ifnum #1\expandafter \@firstoftwo
 \else \expandafter \@secondoftwo
 \fi
}%
\providecommand \@ifx [1]{%
 \ifx #1\expandafter \@firstoftwo
 \else \expandafter \@secondoftwo
 \fi
}%
\providecommand \natexlab [1]{#1}%
\providecommand \enquote  [1]{``#1''}%
\providecommand \bibnamefont  [1]{#1}%
\providecommand \bibfnamefont [1]{#1}%
\providecommand \citenamefont [1]{#1}%
\providecommand \href@noop [0]{\@secondoftwo}%
\providecommand \href [0]{\begingroup \@sanitize@url \@href}%
\providecommand \@href[1]{\@@startlink{#1}\@@href}%
\providecommand \@@href[1]{\endgroup#1\@@endlink}%
\providecommand \@sanitize@url [0]{\catcode `\\12\catcode `\$12\catcode `\&12\catcode `\#12\catcode `\^12\catcode `\_12\catcode `\%12\relax}%
\providecommand \@@startlink[1]{}%
\providecommand \@@endlink[0]{}%
\providecommand \url  [0]{\begingroup\@sanitize@url \@url }%
\providecommand \@url [1]{\endgroup\@href {#1}{\urlprefix }}%
\providecommand \urlprefix  [0]{URL }%
\providecommand \Eprint [0]{\href }%
\providecommand \doibase [0]{https://doi.org/}%
\providecommand \selectlanguage [0]{\@gobble}%
\providecommand \bibinfo  [0]{\@secondoftwo}%
\providecommand \bibfield  [0]{\@secondoftwo}%
\providecommand \translation [1]{[#1]}%
\providecommand \BibitemOpen [0]{}%
\providecommand \bibitemStop [0]{}%
\providecommand \bibitemNoStop [0]{.\EOS\space}%
\providecommand \EOS [0]{\spacefactor3000\relax}%
\providecommand \BibitemShut  [1]{\csname bibitem#1\endcsname}%
\let\auto@bib@innerbib\@empty
\bibitem [{\citenamefont {Hohenberg}\ and\ \citenamefont {Kohn}(1964)}]{Kohn_1964}%
  \BibitemOpen
  \bibfield  {author} {\bibinfo {author} {\bibfnamefont {P.}~\bibnamefont {Hohenberg}}\ and\ \bibinfo {author} {\bibfnamefont {W.}~\bibnamefont {Kohn}},\ }\bibfield  {title} {\bibinfo {title} {Inhomogeneous electron gas},\ }\href {https://doi.org/10.1103/PhysRev.136.B864} {\bibfield  {journal} {\bibinfo  {journal} {Phys. Rev.}\ }\textbf {\bibinfo {volume} {136}},\ \bibinfo {pages} {B864} (\bibinfo {year} {1964})}\BibitemShut {NoStop}%
\bibitem [{\citenamefont {Kohn}\ and\ \citenamefont {Sham}(1965)}]{Kohn_1965}%
  \BibitemOpen
  \bibfield  {author} {\bibinfo {author} {\bibfnamefont {W.}~\bibnamefont {Kohn}}\ and\ \bibinfo {author} {\bibfnamefont {L.~J.}\ \bibnamefont {Sham}},\ }\bibfield  {title} {\bibinfo {title} {Self-consistent equations including exchange and correlation effects},\ }\href {https://doi.org/10.1103/PhysRev.140.A1133} {\bibfield  {journal} {\bibinfo  {journal} {Phys. Rev.}\ }\textbf {\bibinfo {volume} {140}},\ \bibinfo {pages} {A1133} (\bibinfo {year} {1965})}\BibitemShut {NoStop}%
\bibitem [{\citenamefont {Perdew}\ and\ \citenamefont {Kurth}(2003)}]{Perdew_2003}%
  \BibitemOpen
  \bibfield  {author} {\bibinfo {author} {\bibfnamefont {J.~P.}\ \bibnamefont {Perdew}}\ and\ \bibinfo {author} {\bibfnamefont {S.}~\bibnamefont {Kurth}},\ }\bibinfo {title} {Density functionals for non-relativistic coulomb systems in the new century},\ in\ \href {https://doi.org/10.1007/3-540-37072-2_1} {\emph {\bibinfo {booktitle} {A Primer in Density Functional Theory}}},\ \bibinfo {editor} {edited by\ \bibinfo {editor} {\bibfnamefont {C.}~\bibnamefont {Fiolhais}}, \bibinfo {editor} {\bibfnamefont {F.}~\bibnamefont {Nogueira}},\ and\ \bibinfo {editor} {\bibfnamefont {M.~A.~L.}\ \bibnamefont {Marques}}}\ (\bibinfo  {publisher} {Springer Berlin Heidelberg},\ \bibinfo {address} {Berlin, Heidelberg},\ \bibinfo {year} {2003})\ pp.\ \bibinfo {pages} {1--55}\BibitemShut {NoStop}%
\bibitem [{\citenamefont {Laughlin}(1983)}]{Laughlin_1983}%
  \BibitemOpen
  \bibfield  {author} {\bibinfo {author} {\bibfnamefont {R.~B.}\ \bibnamefont {Laughlin}},\ }\bibfield  {title} {\bibinfo {title} {Anomalous quantum hall effect: An incompressible quantum fluid with fractionally charged excitations},\ }\href {https://doi.org/10.1103/PhysRevLett.50.1395} {\bibfield  {journal} {\bibinfo  {journal} {Phys. Rev. Lett.}\ }\textbf {\bibinfo {volume} {50}},\ \bibinfo {pages} {1395} (\bibinfo {year} {1983})}\BibitemShut {NoStop}%
\bibitem [{\citenamefont {Singh}\ \emph {et~al.}(2024)\citenamefont {Singh}, \citenamefont {Wang}, \citenamefont {Tai}, \citenamefont {Calhoun}, \citenamefont {Villegas~Rosales}, \citenamefont {Madathil}, \citenamefont {Gupta}, \citenamefont {Baldwin}, \citenamefont {Pfeiffer},\ and\ \citenamefont {Shayegan}}]{Singh_2024}%
  \BibitemOpen
  \bibfield  {author} {\bibinfo {author} {\bibfnamefont {S.~K.}\ \bibnamefont {Singh}}, \bibinfo {author} {\bibfnamefont {C.}~\bibnamefont {Wang}}, \bibinfo {author} {\bibfnamefont {C.~T.}\ \bibnamefont {Tai}}, \bibinfo {author} {\bibfnamefont {C.~S.}\ \bibnamefont {Calhoun}}, \bibinfo {author} {\bibfnamefont {K.~A.}\ \bibnamefont {Villegas~Rosales}}, \bibinfo {author} {\bibfnamefont {P.~T.}\ \bibnamefont {Madathil}}, \bibinfo {author} {\bibfnamefont {A.}~\bibnamefont {Gupta}}, \bibinfo {author} {\bibfnamefont {K.~W.}\ \bibnamefont {Baldwin}}, \bibinfo {author} {\bibfnamefont {L.~N.}\ \bibnamefont {Pfeiffer}},\ and\ \bibinfo {author} {\bibfnamefont {M.}~\bibnamefont {Shayegan}},\ }\bibfield  {title} {\bibinfo {title} {Topological phase transition between jain states and daughter states of the \ensuremath{\nu}= 1/2 fractional quantum hall state},\ }\href {https://doi.org/10.1038/s41567-024-02517-w} {\bibfield  {journal} {\bibinfo  {journal} {Nature Physics}\ }\textbf {\bibinfo {volume} {20}},\ \bibinfo {pages}
  {1247} (\bibinfo {year} {2024})}\BibitemShut {NoStop}%
\bibitem [{\citenamefont {Singh}\ \emph {et~al.}(2025)\citenamefont {Singh}, \citenamefont {Wang}, \citenamefont {Gupta}, \citenamefont {Baldwin}, \citenamefont {Pfeiffer},\ and\ \citenamefont {Shayegan}}]{Singh_2025}%
  \BibitemOpen
  \bibfield  {author} {\bibinfo {author} {\bibfnamefont {S.~K.}\ \bibnamefont {Singh}}, \bibinfo {author} {\bibfnamefont {C.}~\bibnamefont {Wang}}, \bibinfo {author} {\bibfnamefont {A.}~\bibnamefont {Gupta}}, \bibinfo {author} {\bibfnamefont {K.~W.}\ \bibnamefont {Baldwin}}, \bibinfo {author} {\bibfnamefont {L.~N.}\ \bibnamefont {Pfeiffer}},\ and\ \bibinfo {author} {\bibfnamefont {M.}~\bibnamefont {Shayegan}},\ }\href {https://arxiv.org/abs/2510.03983} {\bibinfo {title} {Fractional quantum hall state at $\nu = 1/2$ with energy gap up to 6 k, and possible transition from one- to two-component state}} (\bibinfo {year} {2025}),\ \Eprint {https://arxiv.org/abs/2510.03983} {arXiv:2510.03983 [cond-mat.mes-hall]} \BibitemShut {NoStop}%
\bibitem [{\citenamefont {Chung}\ \emph {et~al.}(2021)\citenamefont {Chung}, \citenamefont {Villegas~Rosales}, \citenamefont {Baldwin}, \citenamefont {Madathil}, \citenamefont {West}, \citenamefont {Shayegan},\ and\ \citenamefont {Pfeiffer}}]{Chung_2021}%
  \BibitemOpen
  \bibfield  {author} {\bibinfo {author} {\bibfnamefont {Y.~J.}\ \bibnamefont {Chung}}, \bibinfo {author} {\bibfnamefont {K.~A.}\ \bibnamefont {Villegas~Rosales}}, \bibinfo {author} {\bibfnamefont {K.~W.}\ \bibnamefont {Baldwin}}, \bibinfo {author} {\bibfnamefont {P.~T.}\ \bibnamefont {Madathil}}, \bibinfo {author} {\bibfnamefont {K.~W.}\ \bibnamefont {West}}, \bibinfo {author} {\bibfnamefont {M.}~\bibnamefont {Shayegan}},\ and\ \bibinfo {author} {\bibfnamefont {L.~N.}\ \bibnamefont {Pfeiffer}},\ }\bibfield  {title} {\bibinfo {title} {Ultra-high-quality two-dimensional electron systems},\ }\href {https://doi.org/10.1038/s41563-021-00942-3} {\bibfield  {journal} {\bibinfo  {journal} {Nature Materials}\ }\textbf {\bibinfo {volume} {20}},\ \bibinfo {pages} {632} (\bibinfo {year} {2021})}\BibitemShut {NoStop}%
\bibitem [{\citenamefont {Moore}\ and\ \citenamefont {Read}(1991)}]{Moore_1991}%
  \BibitemOpen
  \bibfield  {author} {\bibinfo {author} {\bibfnamefont {G.}~\bibnamefont {Moore}}\ and\ \bibinfo {author} {\bibfnamefont {N.}~\bibnamefont {Read}},\ }\bibfield  {title} {\bibinfo {title} {Nonabelions in the fractional quantum hall effect},\ }\href {https://doi.org/https://doi.org/10.1016/0550-3213(91)90407-O} {\bibfield  {journal} {\bibinfo  {journal} {Nuclear Physics B}\ }\textbf {\bibinfo {volume} {360}},\ \bibinfo {pages} {362} (\bibinfo {year} {1991})}\BibitemShut {NoStop}%
\bibitem [{\citenamefont {Nayak}\ \emph {et~al.}(2008)\citenamefont {Nayak}, \citenamefont {Simon}, \citenamefont {Stern}, \citenamefont {Freedman},\ and\ \citenamefont {Das~Sarma}}]{Nayak_2008}%
  \BibitemOpen
  \bibfield  {author} {\bibinfo {author} {\bibfnamefont {C.}~\bibnamefont {Nayak}}, \bibinfo {author} {\bibfnamefont {S.~H.}\ \bibnamefont {Simon}}, \bibinfo {author} {\bibfnamefont {A.}~\bibnamefont {Stern}}, \bibinfo {author} {\bibfnamefont {M.}~\bibnamefont {Freedman}},\ and\ \bibinfo {author} {\bibfnamefont {S.}~\bibnamefont {Das~Sarma}},\ }\bibfield  {title} {\bibinfo {title} {Non-abelian anyons and topological quantum computation},\ }\href {https://doi.org/10.1103/RevModPhys.80.1083} {\bibfield  {journal} {\bibinfo  {journal} {Rev. Mod. Phys.}\ }\textbf {\bibinfo {volume} {80}},\ \bibinfo {pages} {1083} (\bibinfo {year} {2008})}\BibitemShut {NoStop}%
\bibitem [{\citenamefont {Sharma}\ \emph {et~al.}(2024)\citenamefont {Sharma}, \citenamefont {Balram},\ and\ \citenamefont {Jain}}]{Sharma_2024}%
  \BibitemOpen
  \bibfield  {author} {\bibinfo {author} {\bibfnamefont {A.}~\bibnamefont {Sharma}}, \bibinfo {author} {\bibfnamefont {A.~C.}\ \bibnamefont {Balram}},\ and\ \bibinfo {author} {\bibfnamefont {J.~K.}\ \bibnamefont {Jain}},\ }\bibfield  {title} {\bibinfo {title} {Composite-fermion pairing at half-filled and quarter-filled lowest landau level},\ }\href {https://doi.org/10.1103/PhysRevB.109.035306} {\bibfield  {journal} {\bibinfo  {journal} {Phys. Rev. B}\ }\textbf {\bibinfo {volume} {109}},\ \bibinfo {pages} {035306} (\bibinfo {year} {2024})}\BibitemShut {NoStop}%
\bibitem [{\citenamefont {Drummond}\ and\ \citenamefont {Needs}(2009)}]{Drummond_2009}%
  \BibitemOpen
  \bibfield  {author} {\bibinfo {author} {\bibfnamefont {N.~D.}\ \bibnamefont {Drummond}}\ and\ \bibinfo {author} {\bibfnamefont {R.~J.}\ \bibnamefont {Needs}},\ }\bibfield  {title} {\bibinfo {title} {Phase diagram of the low-density two-dimensional homogeneous electron gas},\ }\href {https://doi.org/10.1103/PhysRevLett.102.126402} {\bibfield  {journal} {\bibinfo  {journal} {Phys. Rev. Lett.}\ }\textbf {\bibinfo {volume} {102}},\ \bibinfo {pages} {126402} (\bibinfo {year} {2009})}\BibitemShut {NoStop}%
\bibitem [{\citenamefont {Smith}\ \emph {et~al.}(2024)\citenamefont {Smith}, \citenamefont {Chen}, \citenamefont {Levy}, \citenamefont {Yang}, \citenamefont {Morales},\ and\ \citenamefont {Zhang}}]{Smith_2024}%
  \BibitemOpen
  \bibfield  {author} {\bibinfo {author} {\bibfnamefont {C.}~\bibnamefont {Smith}}, \bibinfo {author} {\bibfnamefont {Y.}~\bibnamefont {Chen}}, \bibinfo {author} {\bibfnamefont {R.}~\bibnamefont {Levy}}, \bibinfo {author} {\bibfnamefont {Y.}~\bibnamefont {Yang}}, \bibinfo {author} {\bibfnamefont {M.~A.}\ \bibnamefont {Morales}},\ and\ \bibinfo {author} {\bibfnamefont {S.}~\bibnamefont {Zhang}},\ }\bibfield  {title} {\bibinfo {title} {Unified variational approach description of ground-state phases of the two-dimensional electron gas},\ }\href {https://doi.org/10.1103/PhysRevLett.133.266504} {\bibfield  {journal} {\bibinfo  {journal} {Phys. Rev. Lett.}\ }\textbf {\bibinfo {volume} {133}},\ \bibinfo {pages} {266504} (\bibinfo {year} {2024})}\BibitemShut {NoStop}%
\bibitem [{\citenamefont {MacDonald}\ \emph {et~al.}(1990)\citenamefont {MacDonald}, \citenamefont {Platzman},\ and\ \citenamefont {Boebinger}}]{MacDonald_1990}%
  \BibitemOpen
  \bibfield  {author} {\bibinfo {author} {\bibfnamefont {A.~H.}\ \bibnamefont {MacDonald}}, \bibinfo {author} {\bibfnamefont {P.~M.}\ \bibnamefont {Platzman}},\ and\ \bibinfo {author} {\bibfnamefont {G.~S.}\ \bibnamefont {Boebinger}},\ }\bibfield  {title} {\bibinfo {title} {Collapse of integer hall gaps in a double-quantum-well system},\ }\href {https://doi.org/10.1103/PhysRevLett.65.775} {\bibfield  {journal} {\bibinfo  {journal} {Phys. Rev. Lett.}\ }\textbf {\bibinfo {volume} {65}},\ \bibinfo {pages} {775} (\bibinfo {year} {1990})}\BibitemShut {NoStop}%
\bibitem [{\citenamefont {Boebinger}\ \emph {et~al.}(1990)\citenamefont {Boebinger}, \citenamefont {Jiang}, \citenamefont {Pfeiffer},\ and\ \citenamefont {West}}]{Boebinger_1990}%
  \BibitemOpen
  \bibfield  {author} {\bibinfo {author} {\bibfnamefont {G.~S.}\ \bibnamefont {Boebinger}}, \bibinfo {author} {\bibfnamefont {H.~W.}\ \bibnamefont {Jiang}}, \bibinfo {author} {\bibfnamefont {L.~N.}\ \bibnamefont {Pfeiffer}},\ and\ \bibinfo {author} {\bibfnamefont {K.~W.}\ \bibnamefont {West}},\ }\bibfield  {title} {\bibinfo {title} {Magnetic-field-driven destruction of quantum hall states in a double quantum well},\ }\href {https://doi.org/10.1103/PhysRevLett.64.1793} {\bibfield  {journal} {\bibinfo  {journal} {Phys. Rev. Lett.}\ }\textbf {\bibinfo {volume} {64}},\ \bibinfo {pages} {1793} (\bibinfo {year} {1990})}\BibitemShut {NoStop}%
\bibitem [{\citenamefont {Eisenstein}\ \emph {et~al.}(1992)\citenamefont {Eisenstein}, \citenamefont {Boebinger}, \citenamefont {Pfeiffer}, \citenamefont {West},\ and\ \citenamefont {He}}]{Eisenstein_1992}%
  \BibitemOpen
  \bibfield  {author} {\bibinfo {author} {\bibfnamefont {J.~P.}\ \bibnamefont {Eisenstein}}, \bibinfo {author} {\bibfnamefont {G.~S.}\ \bibnamefont {Boebinger}}, \bibinfo {author} {\bibfnamefont {L.~N.}\ \bibnamefont {Pfeiffer}}, \bibinfo {author} {\bibfnamefont {K.~W.}\ \bibnamefont {West}},\ and\ \bibinfo {author} {\bibfnamefont {S.}~\bibnamefont {He}},\ }\bibfield  {title} {\bibinfo {title} {New fractional quantum hall state in double-layer two-dimensional electron systems},\ }\href {https://doi.org/10.1103/PhysRevLett.68.1383} {\bibfield  {journal} {\bibinfo  {journal} {Phys. Rev. Lett.}\ }\textbf {\bibinfo {volume} {68}},\ \bibinfo {pages} {1383} (\bibinfo {year} {1992})}\BibitemShut {NoStop}%
\bibitem [{\citenamefont {Narasimhan}\ and\ \citenamefont {Ho}(1995)}]{Narasimhan_1995}%
  \BibitemOpen
  \bibfield  {author} {\bibinfo {author} {\bibfnamefont {S.}~\bibnamefont {Narasimhan}}\ and\ \bibinfo {author} {\bibfnamefont {T.-L.}\ \bibnamefont {Ho}},\ }\bibfield  {title} {\bibinfo {title} {Wigner-crystal phases in bilayer quantum hall systems},\ }\href {https://doi.org/10.1103/PhysRevB.52.12291} {\bibfield  {journal} {\bibinfo  {journal} {Phys. Rev. B}\ }\textbf {\bibinfo {volume} {52}},\ \bibinfo {pages} {12291} (\bibinfo {year} {1995})}\BibitemShut {NoStop}%
\bibitem [{\citenamefont {Manoharan}\ \emph {et~al.}(1996)\citenamefont {Manoharan}, \citenamefont {Suen}, \citenamefont {Santos},\ and\ \citenamefont {Shayegan}}]{Manoharan_1996}%
  \BibitemOpen
  \bibfield  {author} {\bibinfo {author} {\bibfnamefont {H.~C.}\ \bibnamefont {Manoharan}}, \bibinfo {author} {\bibfnamefont {Y.~W.}\ \bibnamefont {Suen}}, \bibinfo {author} {\bibfnamefont {M.~B.}\ \bibnamefont {Santos}},\ and\ \bibinfo {author} {\bibfnamefont {M.}~\bibnamefont {Shayegan}},\ }\bibfield  {title} {\bibinfo {title} {Evidence for a bilayer quantum wigner solid},\ }\href {https://doi.org/10.1103/PhysRevLett.77.1813} {\bibfield  {journal} {\bibinfo  {journal} {Phys. Rev. Lett.}\ }\textbf {\bibinfo {volume} {77}},\ \bibinfo {pages} {1813} (\bibinfo {year} {1996})}\BibitemShut {NoStop}%
\bibitem [{\citenamefont {Hatke}\ \emph {et~al.}(2015)\citenamefont {Hatke}, \citenamefont {Liu}, \citenamefont {Engel}, \citenamefont {Shayegan}, \citenamefont {Pfeiffer}, \citenamefont {West},\ and\ \citenamefont {Baldwin}}]{Hatke_2015}%
  \BibitemOpen
  \bibfield  {author} {\bibinfo {author} {\bibfnamefont {A.~T.}\ \bibnamefont {Hatke}}, \bibinfo {author} {\bibfnamefont {Y.}~\bibnamefont {Liu}}, \bibinfo {author} {\bibfnamefont {L.~W.}\ \bibnamefont {Engel}}, \bibinfo {author} {\bibfnamefont {M.}~\bibnamefont {Shayegan}}, \bibinfo {author} {\bibfnamefont {L.~N.}\ \bibnamefont {Pfeiffer}}, \bibinfo {author} {\bibfnamefont {K.~W.}\ \bibnamefont {West}},\ and\ \bibinfo {author} {\bibfnamefont {K.~W.}\ \bibnamefont {Baldwin}},\ }\bibfield  {title} {\bibinfo {title} {Microwave spectroscopy of the low-filling-factor bilayer electron solid in a wide quantum well},\ }\href {https://doi.org/10.1038/ncomms8071} {\bibfield  {journal} {\bibinfo  {journal} {Nature Communications}\ }\textbf {\bibinfo {volume} {6}},\ \bibinfo {pages} {7071} (\bibinfo {year} {2015})}\BibitemShut {NoStop}%
\bibitem [{\citenamefont {Suen}\ \emph {et~al.}(1991)\citenamefont {Suen}, \citenamefont {Jo}, \citenamefont {Santos}, \citenamefont {Engel}, \citenamefont {Hwang},\ and\ \citenamefont {Shayegan}}]{Suen_1991}%
  \BibitemOpen
  \bibfield  {author} {\bibinfo {author} {\bibfnamefont {Y.~W.}\ \bibnamefont {Suen}}, \bibinfo {author} {\bibfnamefont {J.}~\bibnamefont {Jo}}, \bibinfo {author} {\bibfnamefont {M.~B.}\ \bibnamefont {Santos}}, \bibinfo {author} {\bibfnamefont {L.~W.}\ \bibnamefont {Engel}}, \bibinfo {author} {\bibfnamefont {S.~W.}\ \bibnamefont {Hwang}},\ and\ \bibinfo {author} {\bibfnamefont {M.}~\bibnamefont {Shayegan}},\ }\bibfield  {title} {\bibinfo {title} {Missing integral quantum hall effect in a wide single quantum well},\ }\href {https://doi.org/10.1103/PhysRevB.44.5947} {\bibfield  {journal} {\bibinfo  {journal} {Phys. Rev. B}\ }\textbf {\bibinfo {volume} {44}},\ \bibinfo {pages} {5947} (\bibinfo {year} {1991})}\BibitemShut {NoStop}%
\bibitem [{\citenamefont {Carrasquilla}\ and\ \citenamefont {Melko}(2017)}]{Carrasquilla_2017}%
  \BibitemOpen
  \bibfield  {author} {\bibinfo {author} {\bibfnamefont {J.}~\bibnamefont {Carrasquilla}}\ and\ \bibinfo {author} {\bibfnamefont {R.~G.}\ \bibnamefont {Melko}},\ }\bibfield  {title} {\bibinfo {title} {Machine learning phases of matter},\ }\href {https://doi.org/10.1038/nphys4035} {\bibfield  {journal} {\bibinfo  {journal} {Nature Physics}\ }\textbf {\bibinfo {volume} {13}},\ \bibinfo {pages} {431} (\bibinfo {year} {2017})}\BibitemShut {NoStop}%
\bibitem [{\citenamefont {Carleo}\ and\ \citenamefont {Troyer}(2017)}]{Carleo_2017}%
  \BibitemOpen
  \bibfield  {author} {\bibinfo {author} {\bibfnamefont {G.}~\bibnamefont {Carleo}}\ and\ \bibinfo {author} {\bibfnamefont {M.}~\bibnamefont {Troyer}},\ }\bibfield  {title} {\bibinfo {title} {Solving the quantum many-body problem with artificial neural networks},\ }\href {https://doi.org/10.1126/science.aag2302} {\bibfield  {journal} {\bibinfo  {journal} {Science}\ }\textbf {\bibinfo {volume} {355}},\ \bibinfo {pages} {602} (\bibinfo {year} {2017})},\ \Eprint {https://arxiv.org/abs/https://www.science.org/doi/pdf/10.1126/science.aag2302} {https://www.science.org/doi/pdf/10.1126/science.aag2302} \BibitemShut {NoStop}%
\bibitem [{\citenamefont {Cassella}\ \emph {et~al.}(2023)\citenamefont {Cassella}, \citenamefont {Sutterud}, \citenamefont {Azadi}, \citenamefont {Drummond}, \citenamefont {Pfau}, \citenamefont {Spencer},\ and\ \citenamefont {Foulkes}}]{Cassella_2023}%
  \BibitemOpen
  \bibfield  {author} {\bibinfo {author} {\bibfnamefont {G.}~\bibnamefont {Cassella}}, \bibinfo {author} {\bibfnamefont {H.}~\bibnamefont {Sutterud}}, \bibinfo {author} {\bibfnamefont {S.}~\bibnamefont {Azadi}}, \bibinfo {author} {\bibfnamefont {N.~D.}\ \bibnamefont {Drummond}}, \bibinfo {author} {\bibfnamefont {D.}~\bibnamefont {Pfau}}, \bibinfo {author} {\bibfnamefont {J.~S.}\ \bibnamefont {Spencer}},\ and\ \bibinfo {author} {\bibfnamefont {W.~M.~C.}\ \bibnamefont {Foulkes}},\ }\bibfield  {title} {\bibinfo {title} {Discovering quantum phase transitions with fermionic neural networks},\ }\href {https://doi.org/10.1103/PhysRevLett.130.036401} {\bibfield  {journal} {\bibinfo  {journal} {Phys. Rev. Lett.}\ }\textbf {\bibinfo {volume} {130}},\ \bibinfo {pages} {036401} (\bibinfo {year} {2023})}\BibitemShut {NoStop}%
\bibitem [{\citenamefont {Li}\ \emph {et~al.}(2022)\citenamefont {Li}, \citenamefont {Li},\ and\ \citenamefont {Chen}}]{Li_2022}%
  \BibitemOpen
  \bibfield  {author} {\bibinfo {author} {\bibfnamefont {X.}~\bibnamefont {Li}}, \bibinfo {author} {\bibfnamefont {Z.}~\bibnamefont {Li}},\ and\ \bibinfo {author} {\bibfnamefont {J.}~\bibnamefont {Chen}},\ }\bibfield  {title} {\bibinfo {title} {Ab initio calculation of real solids via neural network ansatz},\ }\href {https://doi.org/10.1038/s41467-022-35627-1} {\bibfield  {journal} {\bibinfo  {journal} {Nature Communications}\ }\textbf {\bibinfo {volume} {13}},\ \bibinfo {pages} {7895} (\bibinfo {year} {2022})}\BibitemShut {NoStop}%
\bibitem [{\citenamefont {Wilson}\ \emph {et~al.}(2023)\citenamefont {Wilson}, \citenamefont {Moroni}, \citenamefont {Holzmann}, \citenamefont {Gao}, \citenamefont {Wudarski}, \citenamefont {Vegge},\ and\ \citenamefont {Bhowmik}}]{Wilson_2023}%
  \BibitemOpen
  \bibfield  {author} {\bibinfo {author} {\bibfnamefont {M.}~\bibnamefont {Wilson}}, \bibinfo {author} {\bibfnamefont {S.}~\bibnamefont {Moroni}}, \bibinfo {author} {\bibfnamefont {M.}~\bibnamefont {Holzmann}}, \bibinfo {author} {\bibfnamefont {N.}~\bibnamefont {Gao}}, \bibinfo {author} {\bibfnamefont {F.}~\bibnamefont {Wudarski}}, \bibinfo {author} {\bibfnamefont {T.}~\bibnamefont {Vegge}},\ and\ \bibinfo {author} {\bibfnamefont {A.}~\bibnamefont {Bhowmik}},\ }\bibfield  {title} {\bibinfo {title} {Neural network ansatz for periodic wave functions and the homogeneous electron gas},\ }\href {https://doi.org/10.1103/PhysRevB.107.235139} {\bibfield  {journal} {\bibinfo  {journal} {Phys. Rev. B}\ }\textbf {\bibinfo {volume} {107}},\ \bibinfo {pages} {235139} (\bibinfo {year} {2023})}\BibitemShut {NoStop}%
\bibitem [{\citenamefont {Pescia}\ \emph {et~al.}(2024)\citenamefont {Pescia}, \citenamefont {Nys}, \citenamefont {Kim}, \citenamefont {Lovato},\ and\ \citenamefont {Carleo}}]{Pescia_2024}%
  \BibitemOpen
  \bibfield  {author} {\bibinfo {author} {\bibfnamefont {G.}~\bibnamefont {Pescia}}, \bibinfo {author} {\bibfnamefont {J.}~\bibnamefont {Nys}}, \bibinfo {author} {\bibfnamefont {J.}~\bibnamefont {Kim}}, \bibinfo {author} {\bibfnamefont {A.}~\bibnamefont {Lovato}},\ and\ \bibinfo {author} {\bibfnamefont {G.}~\bibnamefont {Carleo}},\ }\bibfield  {title} {\bibinfo {title} {Message-passing neural quantum states for the homogeneous electron gas},\ }\href {https://doi.org/10.1103/PhysRevB.110.035108} {\bibfield  {journal} {\bibinfo  {journal} {Phys. Rev. B}\ }\textbf {\bibinfo {volume} {110}},\ \bibinfo {pages} {035108} (\bibinfo {year} {2024})}\BibitemShut {NoStop}%
\bibitem [{\citenamefont {Geier}\ \emph {et~al.}(2025)\citenamefont {Geier}, \citenamefont {Nazaryan}, \citenamefont {Zaklama},\ and\ \citenamefont {Fu}}]{Geier_2025}%
  \BibitemOpen
  \bibfield  {author} {\bibinfo {author} {\bibfnamefont {M.}~\bibnamefont {Geier}}, \bibinfo {author} {\bibfnamefont {K.}~\bibnamefont {Nazaryan}}, \bibinfo {author} {\bibfnamefont {T.}~\bibnamefont {Zaklama}},\ and\ \bibinfo {author} {\bibfnamefont {L.}~\bibnamefont {Fu}},\ }\href {https://arxiv.org/abs/2502.05383} {\bibinfo {title} {Is attention all you need to solve the correlated electron problem?}} (\bibinfo {year} {2025}),\ \Eprint {https://arxiv.org/abs/2502.05383} {arXiv:2502.05383 [cond-mat.str-el]} \BibitemShut {NoStop}%
\bibitem [{\citenamefont {Teng}\ \emph {et~al.}(2025)\citenamefont {Teng}, \citenamefont {Dai},\ and\ \citenamefont {Fu}}]{Teng_2025}%
  \BibitemOpen
  \bibfield  {author} {\bibinfo {author} {\bibfnamefont {Y.}~\bibnamefont {Teng}}, \bibinfo {author} {\bibfnamefont {D.~D.}\ \bibnamefont {Dai}},\ and\ \bibinfo {author} {\bibfnamefont {L.}~\bibnamefont {Fu}},\ }\bibfield  {title} {\bibinfo {title} {Solving the fractional quantum hall problem with self-attention neural network},\ }\href {https://doi.org/10.1103/PhysRevB.111.205117} {\bibfield  {journal} {\bibinfo  {journal} {Phys. Rev. B}\ }\textbf {\bibinfo {volume} {111}},\ \bibinfo {pages} {205117} (\bibinfo {year} {2025})}\BibitemShut {NoStop}%
\bibitem [{\citenamefont {Nazaryan}\ \emph {et~al.}(2025)\citenamefont {Nazaryan}, \citenamefont {Gaggioli}, \citenamefont {Teng},\ and\ \citenamefont {Fu}}]{Nazaryan_2025}%
  \BibitemOpen
  \bibfield  {author} {\bibinfo {author} {\bibfnamefont {K.}~\bibnamefont {Nazaryan}}, \bibinfo {author} {\bibfnamefont {F.}~\bibnamefont {Gaggioli}}, \bibinfo {author} {\bibfnamefont {Y.}~\bibnamefont {Teng}},\ and\ \bibinfo {author} {\bibfnamefont {L.}~\bibnamefont {Fu}},\ }\href {https://arxiv.org/abs/2507.13322} {\bibinfo {title} {Artificial intelligence for quantum matter: Finding a needle in a haystack}} (\bibinfo {year} {2025}),\ \Eprint {https://arxiv.org/abs/2507.13322} {arXiv:2507.13322 [cond-mat.str-el]} \BibitemShut {NoStop}%
\bibitem [{\citenamefont {Li}\ \emph {et~al.}(2025)\citenamefont {Li}, \citenamefont {Ong}, \citenamefont {Geier}, \citenamefont {Lin},\ and\ \citenamefont {Fu}}]{Li_2025}%
  \BibitemOpen
  \bibfield  {author} {\bibinfo {author} {\bibfnamefont {C.-T.}\ \bibnamefont {Li}}, \bibinfo {author} {\bibfnamefont {T.}~\bibnamefont {Ong}}, \bibinfo {author} {\bibfnamefont {M.}~\bibnamefont {Geier}}, \bibinfo {author} {\bibfnamefont {H.}~\bibnamefont {Lin}},\ and\ \bibinfo {author} {\bibfnamefont {L.}~\bibnamefont {Fu}},\ }\href {https://arxiv.org/abs/2509.03683} {\bibinfo {title} {Attention is all you need to solve chiral superconductivity}} (\bibinfo {year} {2025}),\ \Eprint {https://arxiv.org/abs/2509.03683} {arXiv:2509.03683 [cond-mat.supr-con]} \BibitemShut {NoStop}%
\bibitem [{\citenamefont {Shechtman}\ \emph {et~al.}(1984)\citenamefont {Shechtman}, \citenamefont {Blech}, \citenamefont {Gratias},\ and\ \citenamefont {Cahn}}]{Shechtman_1984}%
  \BibitemOpen
  \bibfield  {author} {\bibinfo {author} {\bibfnamefont {D.}~\bibnamefont {Shechtman}}, \bibinfo {author} {\bibfnamefont {I.}~\bibnamefont {Blech}}, \bibinfo {author} {\bibfnamefont {D.}~\bibnamefont {Gratias}},\ and\ \bibinfo {author} {\bibfnamefont {J.~W.}\ \bibnamefont {Cahn}},\ }\bibfield  {title} {\bibinfo {title} {Metallic phase with long-range orientational order and no translational symmetry},\ }\href {https://doi.org/10.1103/PhysRevLett.53.1951} {\bibfield  {journal} {\bibinfo  {journal} {Phys. Rev. Lett.}\ }\textbf {\bibinfo {volume} {53}},\ \bibinfo {pages} {1951} (\bibinfo {year} {1984})}\BibitemShut {NoStop}%
\bibitem [{\citenamefont {Levine}\ and\ \citenamefont {Steinhardt}(1984)}]{Levine_1984}%
  \BibitemOpen
  \bibfield  {author} {\bibinfo {author} {\bibfnamefont {D.}~\bibnamefont {Levine}}\ and\ \bibinfo {author} {\bibfnamefont {P.~J.}\ \bibnamefont {Steinhardt}},\ }\bibfield  {title} {\bibinfo {title} {Quasicrystals: A new class of ordered structures},\ }\href {https://doi.org/10.1103/PhysRevLett.53.2477} {\bibfield  {journal} {\bibinfo  {journal} {Phys. Rev. Lett.}\ }\textbf {\bibinfo {volume} {53}},\ \bibinfo {pages} {2477} (\bibinfo {year} {1984})}\BibitemShut {NoStop}%
\bibitem [{\citenamefont {Goldman}\ and\ \citenamefont {Kelton}(1993)}]{Goldman_1993_review}%
  \BibitemOpen
  \bibfield  {author} {\bibinfo {author} {\bibfnamefont {A.~I.}\ \bibnamefont {Goldman}}\ and\ \bibinfo {author} {\bibfnamefont {R.~F.}\ \bibnamefont {Kelton}},\ }\bibfield  {title} {\bibinfo {title} {Quasicrystals and crystalline approximants},\ }\href {https://doi.org/10.1103/RevModPhys.65.213} {\bibfield  {journal} {\bibinfo  {journal} {Rev. Mod. Phys.}\ }\textbf {\bibinfo {volume} {65}},\ \bibinfo {pages} {213} (\bibinfo {year} {1993})}\BibitemShut {NoStop}%
\bibitem [{\citenamefont {Smole{\'n}ski}\ \emph {et~al.}(2021)\citenamefont {Smole{\'n}ski}, \citenamefont {Dolgirev}, \citenamefont {Kuhlenkamp}, \citenamefont {Popert}, \citenamefont {Shimazaki}, \citenamefont {Back}, \citenamefont {Lu}, \citenamefont {Kroner}, \citenamefont {Watanabe}, \citenamefont {Taniguchi}, \citenamefont {Esterlis}, \citenamefont {Demler},\ and\ \citenamefont {Imamo{\u g}lu}}]{Smolenski_2021}%
  \BibitemOpen
  \bibfield  {author} {\bibinfo {author} {\bibfnamefont {T.}~\bibnamefont {Smole{\'n}ski}}, \bibinfo {author} {\bibfnamefont {P.~E.}\ \bibnamefont {Dolgirev}}, \bibinfo {author} {\bibfnamefont {C.}~\bibnamefont {Kuhlenkamp}}, \bibinfo {author} {\bibfnamefont {A.}~\bibnamefont {Popert}}, \bibinfo {author} {\bibfnamefont {Y.}~\bibnamefont {Shimazaki}}, \bibinfo {author} {\bibfnamefont {P.}~\bibnamefont {Back}}, \bibinfo {author} {\bibfnamefont {X.}~\bibnamefont {Lu}}, \bibinfo {author} {\bibfnamefont {M.}~\bibnamefont {Kroner}}, \bibinfo {author} {\bibfnamefont {K.}~\bibnamefont {Watanabe}}, \bibinfo {author} {\bibfnamefont {T.}~\bibnamefont {Taniguchi}}, \bibinfo {author} {\bibfnamefont {I.}~\bibnamefont {Esterlis}}, \bibinfo {author} {\bibfnamefont {E.}~\bibnamefont {Demler}},\ and\ \bibinfo {author} {\bibfnamefont {A.}~\bibnamefont {Imamo{\u g}lu}},\ }\bibfield  {title} {\bibinfo {title} {Signatures of wigner crystal of electrons in a monolayer semiconductor},\ }\href
  {https://doi.org/10.1038/s41586-021-03590-4} {\bibfield  {journal} {\bibinfo  {journal} {Nature}\ }\textbf {\bibinfo {volume} {595}},\ \bibinfo {pages} {53} (\bibinfo {year} {2021})}\BibitemShut {NoStop}%
\bibitem [{\citenamefont {Zhou}\ \emph {et~al.}(2021)\citenamefont {Zhou}, \citenamefont {Sung}, \citenamefont {Brutschea}, \citenamefont {Esterlis}, \citenamefont {Wang}, \citenamefont {Scuri}, \citenamefont {Gelly}, \citenamefont {Heo}, \citenamefont {Taniguchi}, \citenamefont {Watanabe}, \citenamefont {Zar{\'a}nd}, \citenamefont {Lukin}, \citenamefont {Kim}, \citenamefont {Demler},\ and\ \citenamefont {Park}}]{Zhou_2021}%
  \BibitemOpen
  \bibfield  {author} {\bibinfo {author} {\bibfnamefont {Y.}~\bibnamefont {Zhou}}, \bibinfo {author} {\bibfnamefont {J.}~\bibnamefont {Sung}}, \bibinfo {author} {\bibfnamefont {E.}~\bibnamefont {Brutschea}}, \bibinfo {author} {\bibfnamefont {I.}~\bibnamefont {Esterlis}}, \bibinfo {author} {\bibfnamefont {Y.}~\bibnamefont {Wang}}, \bibinfo {author} {\bibfnamefont {G.}~\bibnamefont {Scuri}}, \bibinfo {author} {\bibfnamefont {R.~J.}\ \bibnamefont {Gelly}}, \bibinfo {author} {\bibfnamefont {H.}~\bibnamefont {Heo}}, \bibinfo {author} {\bibfnamefont {T.}~\bibnamefont {Taniguchi}}, \bibinfo {author} {\bibfnamefont {K.}~\bibnamefont {Watanabe}}, \bibinfo {author} {\bibfnamefont {G.}~\bibnamefont {Zar{\'a}nd}}, \bibinfo {author} {\bibfnamefont {M.~D.}\ \bibnamefont {Lukin}}, \bibinfo {author} {\bibfnamefont {P.}~\bibnamefont {Kim}}, \bibinfo {author} {\bibfnamefont {E.}~\bibnamefont {Demler}},\ and\ \bibinfo {author} {\bibfnamefont {H.}~\bibnamefont {Park}},\ }\bibfield  {title} {\bibinfo {title} {Bilayer wigner
  crystals in a transition metal dichalcogenide heterostructure},\ }\href {https://doi.org/10.1038/s41586-021-03560-w} {\bibfield  {journal} {\bibinfo  {journal} {Nature}\ }\textbf {\bibinfo {volume} {595}},\ \bibinfo {pages} {48} (\bibinfo {year} {2021})}\BibitemShut {NoStop}%
\bibitem [{\citenamefont {Li}\ \emph {et~al.}(2021)\citenamefont {Li}, \citenamefont {Li}, \citenamefont {Regan}, \citenamefont {Wang}, \citenamefont {Zhao}, \citenamefont {Kahn}, \citenamefont {Yumigeta}, \citenamefont {Blei}, \citenamefont {Taniguchi}, \citenamefont {Watanabe}, \citenamefont {Tongay}, \citenamefont {Zettl}, \citenamefont {Crommie},\ and\ \citenamefont {Wang}}]{Li_2021}%
  \BibitemOpen
  \bibfield  {author} {\bibinfo {author} {\bibfnamefont {H.}~\bibnamefont {Li}}, \bibinfo {author} {\bibfnamefont {S.}~\bibnamefont {Li}}, \bibinfo {author} {\bibfnamefont {E.~C.}\ \bibnamefont {Regan}}, \bibinfo {author} {\bibfnamefont {D.}~\bibnamefont {Wang}}, \bibinfo {author} {\bibfnamefont {W.}~\bibnamefont {Zhao}}, \bibinfo {author} {\bibfnamefont {S.}~\bibnamefont {Kahn}}, \bibinfo {author} {\bibfnamefont {K.}~\bibnamefont {Yumigeta}}, \bibinfo {author} {\bibfnamefont {M.}~\bibnamefont {Blei}}, \bibinfo {author} {\bibfnamefont {T.}~\bibnamefont {Taniguchi}}, \bibinfo {author} {\bibfnamefont {K.}~\bibnamefont {Watanabe}}, \bibinfo {author} {\bibfnamefont {S.}~\bibnamefont {Tongay}}, \bibinfo {author} {\bibfnamefont {A.}~\bibnamefont {Zettl}}, \bibinfo {author} {\bibfnamefont {M.~F.}\ \bibnamefont {Crommie}},\ and\ \bibinfo {author} {\bibfnamefont {F.}~\bibnamefont {Wang}},\ }\bibfield  {title} {\bibinfo {title} {Imaging two-dimensional generalized wigner crystals},\ }\href
  {https://doi.org/10.1038/s41586-021-03874-9} {\bibfield  {journal} {\bibinfo  {journal} {Nature}\ }\textbf {\bibinfo {volume} {597}},\ \bibinfo {pages} {650} (\bibinfo {year} {2021})}\BibitemShut {NoStop}%
\bibitem [{\citenamefont {Li}\ \emph {et~al.}(2024)\citenamefont {Li}, \citenamefont {Xiang}, \citenamefont {Reddy}, \citenamefont {Devakul}, \citenamefont {Sailus}, \citenamefont {Banerjee}, \citenamefont {Taniguchi}, \citenamefont {Watanabe}, \citenamefont {Tongay}, \citenamefont {Zettl}, \citenamefont {Fu}, \citenamefont {Crommie},\ and\ \citenamefont {Wang}}]{Li_2024}%
  \BibitemOpen
  \bibfield  {author} {\bibinfo {author} {\bibfnamefont {H.}~\bibnamefont {Li}}, \bibinfo {author} {\bibfnamefont {Z.}~\bibnamefont {Xiang}}, \bibinfo {author} {\bibfnamefont {A.~P.}\ \bibnamefont {Reddy}}, \bibinfo {author} {\bibfnamefont {T.}~\bibnamefont {Devakul}}, \bibinfo {author} {\bibfnamefont {R.}~\bibnamefont {Sailus}}, \bibinfo {author} {\bibfnamefont {R.}~\bibnamefont {Banerjee}}, \bibinfo {author} {\bibfnamefont {T.}~\bibnamefont {Taniguchi}}, \bibinfo {author} {\bibfnamefont {K.}~\bibnamefont {Watanabe}}, \bibinfo {author} {\bibfnamefont {S.}~\bibnamefont {Tongay}}, \bibinfo {author} {\bibfnamefont {A.}~\bibnamefont {Zettl}}, \bibinfo {author} {\bibfnamefont {L.}~\bibnamefont {Fu}}, \bibinfo {author} {\bibfnamefont {M.~F.}\ \bibnamefont {Crommie}},\ and\ \bibinfo {author} {\bibfnamefont {F.}~\bibnamefont {Wang}},\ }\bibfield  {title} {\bibinfo {title} {Wigner molecular crystals from multielectron moir{\'e} artificial atoms},\ }\href {https://doi.org/10.1126/science.adk1348} {\bibfield  {journal}
  {\bibinfo  {journal} {Science}\ }\textbf {\bibinfo {volume} {385}},\ \bibinfo {pages} {86} (\bibinfo {year} {2024})},\ \Eprint {https://arxiv.org/abs/https://www.science.org/doi/pdf/10.1126/science.adk1348} {https://www.science.org/doi/pdf/10.1126/science.adk1348} \BibitemShut {NoStop}%
\bibitem [{sup()}]{supplementary}%
  \BibitemOpen
  \href@noop {} {}\bibinfo {note} {See Supplementary Material at url ... .}\BibitemShut {Stop}%
\bibitem [{\citenamefont {von Glehn}\ \emph {et~al.}(2023)\citenamefont {von Glehn}, \citenamefont {Spencer},\ and\ \citenamefont {Pfau}}]{vonGlehn_2023}%
  \BibitemOpen
  \bibfield  {author} {\bibinfo {author} {\bibfnamefont {I.}~\bibnamefont {von Glehn}}, \bibinfo {author} {\bibfnamefont {J.~S.}\ \bibnamefont {Spencer}},\ and\ \bibinfo {author} {\bibfnamefont {D.}~\bibnamefont {Pfau}},\ }\href {https://arxiv.org/abs/2211.13672} {\bibinfo {title} {A self-attention ansatz for ab-initio quantum chemistry}} (\bibinfo {year} {2023}),\ \Eprint {https://arxiv.org/abs/2211.13672} {arXiv:2211.13672 [physics.chem-ph]} \BibitemShut {NoStop}%
\bibitem [{\citenamefont {Foster}\ \emph {et~al.}(2025)\citenamefont {Foster}, \citenamefont {Schätzle}, \citenamefont {Szabó}, \citenamefont {Cheng}, \citenamefont {Köhler}, \citenamefont {Cassella}, \citenamefont {Gao}, \citenamefont {Li}, \citenamefont {Noé},\ and\ \citenamefont {Hermann}}]{Foster_2025}%
  \BibitemOpen
  \bibfield  {author} {\bibinfo {author} {\bibfnamefont {A.}~\bibnamefont {Foster}}, \bibinfo {author} {\bibfnamefont {Z.}~\bibnamefont {Schätzle}}, \bibinfo {author} {\bibfnamefont {P.~B.}\ \bibnamefont {Szabó}}, \bibinfo {author} {\bibfnamefont {L.}~\bibnamefont {Cheng}}, \bibinfo {author} {\bibfnamefont {J.}~\bibnamefont {Köhler}}, \bibinfo {author} {\bibfnamefont {G.}~\bibnamefont {Cassella}}, \bibinfo {author} {\bibfnamefont {N.}~\bibnamefont {Gao}}, \bibinfo {author} {\bibfnamefont {J.}~\bibnamefont {Li}}, \bibinfo {author} {\bibfnamefont {F.}~\bibnamefont {Noé}},\ and\ \bibinfo {author} {\bibfnamefont {J.}~\bibnamefont {Hermann}},\ }\href {https://arxiv.org/abs/2506.19960} {\bibinfo {title} {An ab initio foundation model of wavefunctions that accurately describes chemical bond breaking}} (\bibinfo {year} {2025}),\ \Eprint {https://arxiv.org/abs/2506.19960} {arXiv:2506.19960 [physics.chem-ph]} \BibitemShut {NoStop}%
\bibitem [{\citenamefont {Gaggioli}\ \emph {et~al.}(2026)\citenamefont {Gaggioli}, \citenamefont {Azadi},\ and\ \citenamefont {Fu}}]{Gaggioli_2026}%
  \BibitemOpen
  \bibfield  {author} {\bibinfo {author} {\bibfnamefont {F.}~\bibnamefont {Gaggioli}}, \bibinfo {author} {\bibfnamefont {S.}~\bibnamefont {Azadi}},\ and\ \bibinfo {author} {\bibfnamefont {L.}~\bibnamefont {Fu}},\ }\href {https://arxiv.org/abs/2607.08616} {\bibinfo {title} {Accurate self-attention wavefunctions at large scale}} (\bibinfo {year} {2026}),\ \Eprint {https://arxiv.org/abs/2607.08616} {arXiv:2607.08616 [cond-mat.str-el]} \BibitemShut {NoStop}%
\bibitem [{\citenamefont {Martens}\ and\ \citenamefont {Grosse}(2015)}]{martens2015optimizing}%
  \BibitemOpen
  \bibfield  {author} {\bibinfo {author} {\bibfnamefont {J.}~\bibnamefont {Martens}}\ and\ \bibinfo {author} {\bibfnamefont {R.}~\bibnamefont {Grosse}},\ }\bibfield  {title} {\bibinfo {title} {Optimizing neural networks with kronecker-factored approximate curvature}\ }(\bibinfo  {publisher} {PMLR},\ \bibinfo {year} {2015})\ pp.\ \bibinfo {pages} {2408--2417}\BibitemShut {NoStop}%
\bibitem [{\citenamefont {Jungwirth}\ \emph {et~al.}(1997)\citenamefont {Jungwirth}, \citenamefont {Lay}, \citenamefont {Smr\ifmmode~\check{c}\else \v{c}\fi{}ka},\ and\ \citenamefont {Shayegan}}]{Jungwirth_1997}%
  \BibitemOpen
  \bibfield  {author} {\bibinfo {author} {\bibfnamefont {T.}~\bibnamefont {Jungwirth}}, \bibinfo {author} {\bibfnamefont {T.~S.}\ \bibnamefont {Lay}}, \bibinfo {author} {\bibfnamefont {L.}~\bibnamefont {Smr\ifmmode~\check{c}\else \v{c}\fi{}ka}},\ and\ \bibinfo {author} {\bibfnamefont {M.}~\bibnamefont {Shayegan}},\ }\bibfield  {title} {\bibinfo {title} {Resistance oscillation in wide single quantum wells subject to in-plane magnetic fields},\ }\href {https://doi.org/10.1103/PhysRevB.56.1029} {\bibfield  {journal} {\bibinfo  {journal} {Phys. Rev. B}\ }\textbf {\bibinfo {volume} {56}},\ \bibinfo {pages} {1029} (\bibinfo {year} {1997})}\BibitemShut {NoStop}%
\bibitem [{\citenamefont {Abolfath}\ \emph {et~al.}(1997)\citenamefont {Abolfath}, \citenamefont {Belkhir},\ and\ \citenamefont {Nafari}}]{Abolfath_1997}%
  \BibitemOpen
  \bibfield  {author} {\bibinfo {author} {\bibfnamefont {M.}~\bibnamefont {Abolfath}}, \bibinfo {author} {\bibfnamefont {L.}~\bibnamefont {Belkhir}},\ and\ \bibinfo {author} {\bibfnamefont {N.}~\bibnamefont {Nafari}},\ }\bibfield  {title} {\bibinfo {title} {Quantum hall effect in single wide quantum wells},\ }\href {https://doi.org/10.1103/PhysRevB.55.10643} {\bibfield  {journal} {\bibinfo  {journal} {Phys. Rev. B}\ }\textbf {\bibinfo {volume} {55}},\ \bibinfo {pages} {10643} (\bibinfo {year} {1997})}\BibitemShut {NoStop}%
\bibitem [{\citenamefont {Zhang}\ and\ \citenamefont {Das~Sarma}(1986)}]{Zhang_1986}%
  \BibitemOpen
  \bibfield  {author} {\bibinfo {author} {\bibfnamefont {F.~C.}\ \bibnamefont {Zhang}}\ and\ \bibinfo {author} {\bibfnamefont {S.}~\bibnamefont {Das~Sarma}},\ }\bibfield  {title} {\bibinfo {title} {Excitation gap in the fractional quantum hall effect: Finite layer thickness corrections},\ }\href {https://doi.org/10.1103/PhysRevB.33.2903} {\bibfield  {journal} {\bibinfo  {journal} {Phys. Rev. B}\ }\textbf {\bibinfo {volume} {33}},\ \bibinfo {pages} {2903} (\bibinfo {year} {1986})}\BibitemShut {NoStop}%
\bibitem [{\citenamefont {Goldoni}\ and\ \citenamefont {Peeters}(1996)}]{Goldoni_1996}%
  \BibitemOpen
  \bibfield  {author} {\bibinfo {author} {\bibfnamefont {G.}~\bibnamefont {Goldoni}}\ and\ \bibinfo {author} {\bibfnamefont {F.~M.}\ \bibnamefont {Peeters}},\ }\bibfield  {title} {\bibinfo {title} {Stability, dynamical properties, and melting of a classical bilayer wigner crystal},\ }\href {https://doi.org/10.1103/PhysRevB.53.4591} {\bibfield  {journal} {\bibinfo  {journal} {Phys. Rev. B}\ }\textbf {\bibinfo {volume} {53}},\ \bibinfo {pages} {4591} (\bibinfo {year} {1996})}\BibitemShut {NoStop}%
\bibitem [{\citenamefont {Goldoni}\ and\ \citenamefont {Peeters}(1997)}]{WignerQuantumBilayer}%
  \BibitemOpen
  \bibfield  {author} {\bibinfo {author} {\bibfnamefont {G.}~\bibnamefont {Goldoni}}\ and\ \bibinfo {author} {\bibfnamefont {F.~M.}\ \bibnamefont {Peeters}},\ }\bibfield  {title} {\bibinfo {title} {Wigner crystallization in quantum electron bilayers},\ }\href {https://doi.org/10.1209/epl/i1997-00544-3} {\bibfield  {journal} {\bibinfo  {journal} {Europhysics Letters}\ }\textbf {\bibinfo {volume} {37}},\ \bibinfo {pages} {293} (\bibinfo {year} {1997})}\BibitemShut {NoStop}%
\bibitem [{\citenamefont {Pleasants}\ \emph {et~al.}(1996)\citenamefont {Pleasants}, \citenamefont {Baake},\ and\ \citenamefont {Roth}}]{Pleasants_1996}%
  \BibitemOpen
  \bibfield  {author} {\bibinfo {author} {\bibfnamefont {P.~A.~B.}\ \bibnamefont {Pleasants}}, \bibinfo {author} {\bibfnamefont {M.}~\bibnamefont {Baake}},\ and\ \bibinfo {author} {\bibfnamefont {J.}~\bibnamefont {Roth}},\ }\bibfield  {title} {\bibinfo {title} {Planar coincidences for n-fold symmetry},\ }\href {https://doi.org/10.1063/1.531424} {\bibfield  {journal} {\bibinfo  {journal} {J. Math. Phys.}\ }\textbf {\bibinfo {volume} {37}},\ \bibinfo {pages} {1029} (\bibinfo {year} {1996})}\BibitemShut {NoStop}%
\bibitem [{\citenamefont {Mele}(2010)}]{Mele_2010}%
  \BibitemOpen
  \bibfield  {author} {\bibinfo {author} {\bibfnamefont {E.~J.}\ \bibnamefont {Mele}},\ }\bibfield  {title} {\bibinfo {title} {Commensuration and interlayer coherence in twisted bilayer graphene},\ }\href {https://doi.org/10.1103/PhysRevB.81.161405} {\bibfield  {journal} {\bibinfo  {journal} {Phys. Rev. B}\ }\textbf {\bibinfo {volume} {81}},\ \bibinfo {pages} {161405} (\bibinfo {year} {2010})}\BibitemShut {NoStop}%
\bibitem [{\citenamefont {Ahn}\ \emph {et~al.}(2018)\citenamefont {Ahn}, \citenamefont {Moon}, \citenamefont {Kim}, \citenamefont {Kim}, \citenamefont {Shin}, \citenamefont {Kim}, \citenamefont {Cha}, \citenamefont {Kahng}, \citenamefont {Kim}, \citenamefont {Koshino}, \citenamefont {Son}, \citenamefont {Yang},\ and\ \citenamefont {Ahn}}]{Ahn_2018}%
  \BibitemOpen
  \bibfield  {author} {\bibinfo {author} {\bibfnamefont {S.~J.}\ \bibnamefont {Ahn}}, \bibinfo {author} {\bibfnamefont {P.}~\bibnamefont {Moon}}, \bibinfo {author} {\bibfnamefont {T.-H.}\ \bibnamefont {Kim}}, \bibinfo {author} {\bibfnamefont {H.-W.}\ \bibnamefont {Kim}}, \bibinfo {author} {\bibfnamefont {H.-C.}\ \bibnamefont {Shin}}, \bibinfo {author} {\bibfnamefont {E.~H.}\ \bibnamefont {Kim}}, \bibinfo {author} {\bibfnamefont {H.~W.}\ \bibnamefont {Cha}}, \bibinfo {author} {\bibfnamefont {S.-J.}\ \bibnamefont {Kahng}}, \bibinfo {author} {\bibfnamefont {P.}~\bibnamefont {Kim}}, \bibinfo {author} {\bibfnamefont {M.}~\bibnamefont {Koshino}}, \bibinfo {author} {\bibfnamefont {Y.-W.}\ \bibnamefont {Son}}, \bibinfo {author} {\bibfnamefont {C.-W.}\ \bibnamefont {Yang}},\ and\ \bibinfo {author} {\bibfnamefont {J.~R.}\ \bibnamefont {Ahn}},\ }\bibfield  {title} {\bibinfo {title} {Dirac electrons in a dodecagonal graphene quasicrystal},\ }\href {https://doi.org/10.1126/science.aar8412} {\bibfield  {journal} {\bibinfo
  {journal} {Science}\ }\textbf {\bibinfo {volume} {361}},\ \bibinfo {pages} {782} (\bibinfo {year} {2018})},\ \Eprint {https://arxiv.org/abs/https://www.science.org/doi/pdf/10.1126/science.aar8412} {https://www.science.org/doi/pdf/10.1126/science.aar8412} \BibitemShut {NoStop}%
\bibitem [{\citenamefont {Yu}\ \emph {et~al.}(2019)\citenamefont {Yu}, \citenamefont {Wu}, \citenamefont {Zhan}, \citenamefont {Katsnelson},\ and\ \citenamefont {Yuan}}]{Yu_2019}%
  \BibitemOpen
  \bibfield  {author} {\bibinfo {author} {\bibfnamefont {G.}~\bibnamefont {Yu}}, \bibinfo {author} {\bibfnamefont {Z.}~\bibnamefont {Wu}}, \bibinfo {author} {\bibfnamefont {Z.}~\bibnamefont {Zhan}}, \bibinfo {author} {\bibfnamefont {M.~I.}\ \bibnamefont {Katsnelson}},\ and\ \bibinfo {author} {\bibfnamefont {S.}~\bibnamefont {Yuan}},\ }\bibfield  {title} {\bibinfo {title} {Dodecagonal bilayer graphene quasicrystal and its approximants},\ }\href {https://doi.org/10.1038/s41524-019-0258-0} {\bibfield  {journal} {\bibinfo  {journal} {npj Computational Materials}\ }\textbf {\bibinfo {volume} {5}},\ \bibinfo {pages} {122} (\bibinfo {year} {2019})}\BibitemShut {NoStop}%
\bibitem [{\citenamefont {Moon}\ \emph {et~al.}(2019)\citenamefont {Moon}, \citenamefont {Koshino},\ and\ \citenamefont {Son}}]{Moon_2019}%
  \BibitemOpen
  \bibfield  {author} {\bibinfo {author} {\bibfnamefont {P.}~\bibnamefont {Moon}}, \bibinfo {author} {\bibfnamefont {M.}~\bibnamefont {Koshino}},\ and\ \bibinfo {author} {\bibfnamefont {Y.-W.}\ \bibnamefont {Son}},\ }\bibfield  {title} {\bibinfo {title} {Quasicrystalline electronic states in ${30}^{\ensuremath{\circ}}$ rotated twisted bilayer graphene},\ }\href {https://doi.org/10.1103/PhysRevB.99.165430} {\bibfield  {journal} {\bibinfo  {journal} {Phys. Rev. B}\ }\textbf {\bibinfo {volume} {99}},\ \bibinfo {pages} {165430} (\bibinfo {year} {2019})}\BibitemShut {NoStop}%
\bibitem [{\citenamefont {Uri}\ \emph {et~al.}(2023)\citenamefont {Uri}, \citenamefont {de~la Barrera}, \citenamefont {Randeria}, \citenamefont {Rodan-Legrain}, \citenamefont {Devakul}, \citenamefont {Crowley}, \citenamefont {Paul}, \citenamefont {Watanabe}, \citenamefont {Taniguchi}, \citenamefont {Lifshitz}, \citenamefont {Fu}, \citenamefont {Ashoori},\ and\ \citenamefont {Jarillo-Herrero}}]{Uri_2023}%
  \BibitemOpen
  \bibfield  {author} {\bibinfo {author} {\bibfnamefont {A.}~\bibnamefont {Uri}}, \bibinfo {author} {\bibfnamefont {S.~C.}\ \bibnamefont {de~la Barrera}}, \bibinfo {author} {\bibfnamefont {M.~T.}\ \bibnamefont {Randeria}}, \bibinfo {author} {\bibfnamefont {D.}~\bibnamefont {Rodan-Legrain}}, \bibinfo {author} {\bibfnamefont {T.}~\bibnamefont {Devakul}}, \bibinfo {author} {\bibfnamefont {P.~J.~D.}\ \bibnamefont {Crowley}}, \bibinfo {author} {\bibfnamefont {N.}~\bibnamefont {Paul}}, \bibinfo {author} {\bibfnamefont {K.}~\bibnamefont {Watanabe}}, \bibinfo {author} {\bibfnamefont {T.}~\bibnamefont {Taniguchi}}, \bibinfo {author} {\bibfnamefont {R.}~\bibnamefont {Lifshitz}}, \bibinfo {author} {\bibfnamefont {L.}~\bibnamefont {Fu}}, \bibinfo {author} {\bibfnamefont {R.~C.}\ \bibnamefont {Ashoori}},\ and\ \bibinfo {author} {\bibfnamefont {P.}~\bibnamefont {Jarillo-Herrero}},\ }\bibfield  {title} {\bibinfo {title} {Superconductivity and strong interactions in a tunable moir{\'e}quasicrystal},\ }\href
  {https://doi.org/10.1038/s41586-023-06294-z} {\bibfield  {journal} {\bibinfo  {journal} {Nature}\ }\textbf {\bibinfo {volume} {620}},\ \bibinfo {pages} {762} (\bibinfo {year} {2023})}\BibitemShut {NoStop}%
\bibitem [{\citenamefont {Xia}\ \emph {et~al.}(2025)\citenamefont {Xia}, \citenamefont {Uri}, \citenamefont {Yan}, \citenamefont {Sharpe}, \citenamefont {Gaggioli}, \citenamefont {Ticea}, \citenamefont {May-Mann}, \citenamefont {Watanabe}, \citenamefont {Taniguchi}, \citenamefont {Fu}, \citenamefont {Devakul}, \citenamefont {Smet},\ and\ \citenamefont {Jarillo-Herrero}}]{Xia_2025}%
  \BibitemOpen
  \bibfield  {author} {\bibinfo {author} {\bibfnamefont {L.-Q.}\ \bibnamefont {Xia}}, \bibinfo {author} {\bibfnamefont {A.}~\bibnamefont {Uri}}, \bibinfo {author} {\bibfnamefont {J.}~\bibnamefont {Yan}}, \bibinfo {author} {\bibfnamefont {A.}~\bibnamefont {Sharpe}}, \bibinfo {author} {\bibfnamefont {F.}~\bibnamefont {Gaggioli}}, \bibinfo {author} {\bibfnamefont {N.~S.}\ \bibnamefont {Ticea}}, \bibinfo {author} {\bibfnamefont {J.}~\bibnamefont {May-Mann}}, \bibinfo {author} {\bibfnamefont {K.}~\bibnamefont {Watanabe}}, \bibinfo {author} {\bibfnamefont {T.}~\bibnamefont {Taniguchi}}, \bibinfo {author} {\bibfnamefont {L.}~\bibnamefont {Fu}}, \bibinfo {author} {\bibfnamefont {T.}~\bibnamefont {Devakul}}, \bibinfo {author} {\bibfnamefont {J.~H.}\ \bibnamefont {Smet}},\ and\ \bibinfo {author} {\bibfnamefont {P.}~\bibnamefont {Jarillo-Herrero}},\ }\href {https://arxiv.org/abs/2509.03583} {\bibinfo {title} {Magic continuum in multi-moir\'e twisted trilayer graphene}} (\bibinfo {year} {2025}),\ \Eprint
  {https://arxiv.org/abs/2509.03583} {arXiv:2509.03583 [cond-mat.mes-hall]} \BibitemShut {NoStop}%
\bibitem [{\citenamefont {Kivelson}\ \emph {et~al.}(1986)\citenamefont {Kivelson}, \citenamefont {Kallin}, \citenamefont {Arovas},\ and\ \citenamefont {Schrieffer}}]{Kivelson_1986}%
  \BibitemOpen
  \bibfield  {author} {\bibinfo {author} {\bibfnamefont {S.}~\bibnamefont {Kivelson}}, \bibinfo {author} {\bibfnamefont {C.}~\bibnamefont {Kallin}}, \bibinfo {author} {\bibfnamefont {D.~P.}\ \bibnamefont {Arovas}},\ and\ \bibinfo {author} {\bibfnamefont {J.~R.}\ \bibnamefont {Schrieffer}},\ }\bibfield  {title} {\bibinfo {title} {Cooperative ring exchange theory of the fractional quantized hall effect},\ }\href {https://doi.org/10.1103/PhysRevLett.56.873} {\bibfield  {journal} {\bibinfo  {journal} {Phys. Rev. Lett.}\ }\textbf {\bibinfo {volume} {56}},\ \bibinfo {pages} {873} (\bibinfo {year} {1986})}\BibitemShut {NoStop}%
\bibitem [{\citenamefont {Sagi}\ and\ \citenamefont {Nussinov}(2016)}]{Sagi_2016}%
  \BibitemOpen
  \bibfield  {author} {\bibinfo {author} {\bibfnamefont {E.}~\bibnamefont {Sagi}}\ and\ \bibinfo {author} {\bibfnamefont {Z.}~\bibnamefont {Nussinov}},\ }\bibfield  {title} {\bibinfo {title} {Emergent quasicrystals in strongly correlated systems},\ }\href {https://doi.org/10.1103/PhysRevB.94.035131} {\bibfield  {journal} {\bibinfo  {journal} {Phys. Rev. B}\ }\textbf {\bibinfo {volume} {94}},\ \bibinfo {pages} {035131} (\bibinfo {year} {2016})}\BibitemShut {NoStop}%
\bibitem [{\citenamefont {Kim}\ \emph {et~al.}(2022)\citenamefont {Kim}, \citenamefont {Murthy}, \citenamefont {Pandey},\ and\ \citenamefont {Kivelson}}]{Kim_2022}%
  \BibitemOpen
  \bibfield  {author} {\bibinfo {author} {\bibfnamefont {K.-S.}\ \bibnamefont {Kim}}, \bibinfo {author} {\bibfnamefont {C.}~\bibnamefont {Murthy}}, \bibinfo {author} {\bibfnamefont {A.}~\bibnamefont {Pandey}},\ and\ \bibinfo {author} {\bibfnamefont {S.~A.}\ \bibnamefont {Kivelson}},\ }\bibfield  {title} {\bibinfo {title} {Interstitial-induced ferromagnetism in a two-dimensional wigner crystal},\ }\href {https://doi.org/10.1103/PhysRevLett.129.227202} {\bibfield  {journal} {\bibinfo  {journal} {Phys. Rev. Lett.}\ }\textbf {\bibinfo {volume} {129}},\ \bibinfo {pages} {227202} (\bibinfo {year} {2022})}\BibitemShut {NoStop}%
\bibitem [{\citenamefont {Kalugin}\ \emph {et~al.}(1985)\citenamefont {Kalugin}, \citenamefont {Kitayev},\ and\ \citenamefont {Levitov}}]{Kalugin_1985}%
  \BibitemOpen
  \bibfield  {author} {\bibinfo {author} {\bibfnamefont {P.}~\bibnamefont {Kalugin}}, \bibinfo {author} {\bibfnamefont {A.~Y.}\ \bibnamefont {Kitayev}},\ and\ \bibinfo {author} {\bibfnamefont {L.}~\bibnamefont {Levitov}},\ }\bibfield  {title} {\bibinfo {title} {{6-dimensional properties of Al0.86Mn0.14 alloy}},\ }\href {https://doi.org/10.1051/jphyslet:019850046013060100} {\bibfield  {journal} {\bibinfo  {journal} {{Journal de Physique Lettres}}\ }\textbf {\bibinfo {volume} {46}},\ \bibinfo {pages} {601} (\bibinfo {year} {1985})}\BibitemShut {NoStop}%
\bibitem [{\citenamefont {Nuckolls}\ \emph {et~al.}(2025)\citenamefont {Nuckolls}, \citenamefont {Paul}, \citenamefont {Chen}, \citenamefont {Gaggioli}, \citenamefont {Wakefield}, \citenamefont {Auslender}, \citenamefont {Gardener}, \citenamefont {Akey}, \citenamefont {Graf}, \citenamefont {Suzuki}, \citenamefont {Bell}, \citenamefont {Fu},\ and\ \citenamefont {Checkelsky}}]{Nuckolls_2025}%
  \BibitemOpen
  \bibfield  {author} {\bibinfo {author} {\bibfnamefont {K.~P.}\ \bibnamefont {Nuckolls}}, \bibinfo {author} {\bibfnamefont {N.}~\bibnamefont {Paul}}, \bibinfo {author} {\bibfnamefont {A.}~\bibnamefont {Chen}}, \bibinfo {author} {\bibfnamefont {F.}~\bibnamefont {Gaggioli}}, \bibinfo {author} {\bibfnamefont {J.~P.}\ \bibnamefont {Wakefield}}, \bibinfo {author} {\bibfnamefont {A.}~\bibnamefont {Auslender}}, \bibinfo {author} {\bibfnamefont {J.}~\bibnamefont {Gardener}}, \bibinfo {author} {\bibfnamefont {A.~J.}\ \bibnamefont {Akey}}, \bibinfo {author} {\bibfnamefont {D.}~\bibnamefont {Graf}}, \bibinfo {author} {\bibfnamefont {T.}~\bibnamefont {Suzuki}}, \bibinfo {author} {\bibfnamefont {D.~C.}\ \bibnamefont {Bell}}, \bibinfo {author} {\bibfnamefont {L.}~\bibnamefont {Fu}},\ and\ \bibinfo {author} {\bibfnamefont {J.~G.}\ \bibnamefont {Checkelsky}},\ }\href {https://arxiv.org/abs/2510.26880} {\bibinfo {title} {Higher-dimensional fermiology in bulk moir\'e metals}} (\bibinfo {year} {2025}),\ \Eprint
  {https://arxiv.org/abs/2510.26880} {arXiv:2510.26880 [cond-mat.mtrl-sci]} \BibitemShut {NoStop}%
\bibitem [{\citenamefont {Kraus}\ \emph {et~al.}(2013)\citenamefont {Kraus}, \citenamefont {Ringel},\ and\ \citenamefont {Zilberberg}}]{Kraus_2013}%
  \BibitemOpen
  \bibfield  {author} {\bibinfo {author} {\bibfnamefont {Y.~E.}\ \bibnamefont {Kraus}}, \bibinfo {author} {\bibfnamefont {Z.}~\bibnamefont {Ringel}},\ and\ \bibinfo {author} {\bibfnamefont {O.}~\bibnamefont {Zilberberg}},\ }\bibfield  {title} {\bibinfo {title} {Four-dimensional quantum hall effect in a two-dimensional quasicrystal},\ }\href {https://doi.org/10.1103/PhysRevLett.111.226401} {\bibfield  {journal} {\bibinfo  {journal} {Phys. Rev. Lett.}\ }\textbf {\bibinfo {volume} {111}},\ \bibinfo {pages} {226401} (\bibinfo {year} {2013})}\BibitemShut {NoStop}%
\bibitem [{\citenamefont {Lohse}\ \emph {et~al.}(2018)\citenamefont {Lohse}, \citenamefont {Schweizer}, \citenamefont {Price}, \citenamefont {Zilberberg},\ and\ \citenamefont {Bloch}}]{Lohse_2018}%
  \BibitemOpen
  \bibfield  {author} {\bibinfo {author} {\bibfnamefont {M.}~\bibnamefont {Lohse}}, \bibinfo {author} {\bibfnamefont {C.}~\bibnamefont {Schweizer}}, \bibinfo {author} {\bibfnamefont {H.~M.}\ \bibnamefont {Price}}, \bibinfo {author} {\bibfnamefont {O.}~\bibnamefont {Zilberberg}},\ and\ \bibinfo {author} {\bibfnamefont {I.}~\bibnamefont {Bloch}},\ }\bibfield  {title} {\bibinfo {title} {Exploring 4d quantum hall physics with a 2d topological charge pump},\ }\href {https://doi.org/10.1038/nature25000} {\bibfield  {journal} {\bibinfo  {journal} {Nature}\ }\textbf {\bibinfo {volume} {553}},\ \bibinfo {pages} {55} (\bibinfo {year} {2018})}\BibitemShut {NoStop}%
\bibitem [{\citenamefont {Yoon}\ \emph {et~al.}(1999)\citenamefont {Yoon}, \citenamefont {Li}, \citenamefont {Shahar}, \citenamefont {Tsui},\ and\ \citenamefont {Shayegan}}]{Yoon_1999}%
  \BibitemOpen
  \bibfield  {author} {\bibinfo {author} {\bibfnamefont {J.}~\bibnamefont {Yoon}}, \bibinfo {author} {\bibfnamefont {C.~C.}\ \bibnamefont {Li}}, \bibinfo {author} {\bibfnamefont {D.}~\bibnamefont {Shahar}}, \bibinfo {author} {\bibfnamefont {D.~C.}\ \bibnamefont {Tsui}},\ and\ \bibinfo {author} {\bibfnamefont {M.}~\bibnamefont {Shayegan}},\ }\bibfield  {title} {\bibinfo {title} {Wigner crystallization and metal-insulator transition of two-dimensional holes in gaas at $\mathit{B}\phantom{\rule{0ex}{0ex}}=\phantom{\rule{0ex}{0ex}}0$},\ }\href {https://doi.org/10.1103/PhysRevLett.82.1744} {\bibfield  {journal} {\bibinfo  {journal} {Phys. Rev. Lett.}\ }\textbf {\bibinfo {volume} {82}},\ \bibinfo {pages} {1744} (\bibinfo {year} {1999})}\BibitemShut {NoStop}%
\bibitem [{\citenamefont {Tsui}\ \emph {et~al.}(2024)\citenamefont {Tsui}, \citenamefont {He}, \citenamefont {Hu}, \citenamefont {Lake}, \citenamefont {Wang}, \citenamefont {Watanabe}, \citenamefont {Taniguchi}, \citenamefont {Zaletel},\ and\ \citenamefont {Yazdani}}]{Tsui_2024}%
  \BibitemOpen
  \bibfield  {author} {\bibinfo {author} {\bibfnamefont {Y.-C.}\ \bibnamefont {Tsui}}, \bibinfo {author} {\bibfnamefont {M.}~\bibnamefont {He}}, \bibinfo {author} {\bibfnamefont {Y.}~\bibnamefont {Hu}}, \bibinfo {author} {\bibfnamefont {E.}~\bibnamefont {Lake}}, \bibinfo {author} {\bibfnamefont {T.}~\bibnamefont {Wang}}, \bibinfo {author} {\bibfnamefont {K.}~\bibnamefont {Watanabe}}, \bibinfo {author} {\bibfnamefont {T.}~\bibnamefont {Taniguchi}}, \bibinfo {author} {\bibfnamefont {M.~P.}\ \bibnamefont {Zaletel}},\ and\ \bibinfo {author} {\bibfnamefont {A.}~\bibnamefont {Yazdani}},\ }\bibfield  {title} {\bibinfo {title} {Direct observation of a magnetic-field-induced wigner crystal},\ }\href {https://doi.org/10.1038/s41586-024-07212-7} {\bibfield  {journal} {\bibinfo  {journal} {Nature}\ }\textbf {\bibinfo {volume} {628}},\ \bibinfo {pages} {287} (\bibinfo {year} {2024})}\BibitemShut {NoStop}%
\bibitem [{\citenamefont {Haji-Akbari}\ \emph {et~al.}(2009)\citenamefont {Haji-Akbari}, \citenamefont {Engel}, \citenamefont {Keys}, \citenamefont {Zheng}, \citenamefont {Petschek}, \citenamefont {Palffy-Muhoray},\ and\ \citenamefont {Glotzer}}]{Haji-Akbari_2009}%
  \BibitemOpen
  \bibfield  {author} {\bibinfo {author} {\bibfnamefont {A.}~\bibnamefont {Haji-Akbari}}, \bibinfo {author} {\bibfnamefont {M.}~\bibnamefont {Engel}}, \bibinfo {author} {\bibfnamefont {A.~S.}\ \bibnamefont {Keys}}, \bibinfo {author} {\bibfnamefont {X.}~\bibnamefont {Zheng}}, \bibinfo {author} {\bibfnamefont {R.~G.}\ \bibnamefont {Petschek}}, \bibinfo {author} {\bibfnamefont {P.}~\bibnamefont {Palffy-Muhoray}},\ and\ \bibinfo {author} {\bibfnamefont {S.~C.}\ \bibnamefont {Glotzer}},\ }\bibfield  {title} {\bibinfo {title} {Disordered, quasicrystalline and crystalline phases of densely packed tetrahedra},\ }\href {https://doi.org/10.1038/nature08641} {\bibfield  {journal} {\bibinfo  {journal} {Nature}\ }\textbf {\bibinfo {volume} {462}},\ \bibinfo {pages} {773} (\bibinfo {year} {2009})}\BibitemShut {NoStop}%
\bibitem [{\citenamefont {Iacovella}\ \emph {et~al.}(2011)\citenamefont {Iacovella}, \citenamefont {Keys},\ and\ \citenamefont {Glotzer}}]{Iacovella_2011}%
  \BibitemOpen
  \bibfield  {author} {\bibinfo {author} {\bibfnamefont {C.~R.}\ \bibnamefont {Iacovella}}, \bibinfo {author} {\bibfnamefont {A.~S.}\ \bibnamefont {Keys}},\ and\ \bibinfo {author} {\bibfnamefont {S.~C.}\ \bibnamefont {Glotzer}},\ }\bibfield  {title} {\bibinfo {title} {Self-assembly of soft-matter quasicrystals and their approximants},\ }\href {https://doi.org/10.1073/pnas.1019763108} {\bibfield  {journal} {\bibinfo  {journal} {Proceedings of the National Academy of Sciences}\ }\textbf {\bibinfo {volume} {108}},\ \bibinfo {pages} {20935} (\bibinfo {year} {2011})},\ \Eprint {https://arxiv.org/abs/https://www.pnas.org/doi/pdf/10.1073/pnas.1019763108} {https://www.pnas.org/doi/pdf/10.1073/pnas.1019763108} \BibitemShut {NoStop}%
\bibitem [{\citenamefont {Fayen}\ \emph {et~al.}(2024)\citenamefont {Fayen}, \citenamefont {Filion}, \citenamefont {Foffi},\ and\ \citenamefont {Smallenburg}}]{Fayen_2024}%
  \BibitemOpen
  \bibfield  {author} {\bibinfo {author} {\bibfnamefont {E.}~\bibnamefont {Fayen}}, \bibinfo {author} {\bibfnamefont {L.}~\bibnamefont {Filion}}, \bibinfo {author} {\bibfnamefont {G.}~\bibnamefont {Foffi}},\ and\ \bibinfo {author} {\bibfnamefont {F.}~\bibnamefont {Smallenburg}},\ }\bibfield  {title} {\bibinfo {title} {Quasicrystal of binary hard spheres on a plane stabilized by configurational entropy},\ }\href {https://doi.org/10.1103/PhysRevLett.132.048202} {\bibfield  {journal} {\bibinfo  {journal} {Phys. Rev. Lett.}\ }\textbf {\bibinfo {volume} {132}},\ \bibinfo {pages} {048202} (\bibinfo {year} {2024})}\BibitemShut {NoStop}%
\bibitem [{\citenamefont {Peterson}\ and\ \citenamefont {Das~Sarma}(2010)}]{Peterson_2010}%
  \BibitemOpen
  \bibfield  {author} {\bibinfo {author} {\bibfnamefont {M.~R.}\ \bibnamefont {Peterson}}\ and\ \bibinfo {author} {\bibfnamefont {S.}~\bibnamefont {Das~Sarma}},\ }\bibfield  {title} {\bibinfo {title} {Quantum hall phase diagram of half-filled bilayers in the lowest and the second orbital landau levels: Abelian versus non-abelian incompressible fractional quantum hall states},\ }\href {https://doi.org/10.1103/PhysRevB.81.165304} {\bibfield  {journal} {\bibinfo  {journal} {Phys. Rev. B}\ }\textbf {\bibinfo {volume} {81}},\ \bibinfo {pages} {165304} (\bibinfo {year} {2010})}\BibitemShut {NoStop}%
\bibitem [{\citenamefont {Thiebaut}\ \emph {et~al.}(2015)\citenamefont {Thiebaut}, \citenamefont {Regnault},\ and\ \citenamefont {Goerbig}}]{Thiebaut_2015}%
  \BibitemOpen
  \bibfield  {author} {\bibinfo {author} {\bibfnamefont {N.}~\bibnamefont {Thiebaut}}, \bibinfo {author} {\bibfnamefont {N.}~\bibnamefont {Regnault}},\ and\ \bibinfo {author} {\bibfnamefont {M.~O.}\ \bibnamefont {Goerbig}},\ }\bibfield  {title} {\bibinfo {title} {Fractional quantum hall states versus wigner crystals in wide quantum wells in the half-filled lowest and second landau levels},\ }\href {https://doi.org/10.1103/PhysRevB.92.245401} {\bibfield  {journal} {\bibinfo  {journal} {Phys. Rev. B}\ }\textbf {\bibinfo {volume} {92}},\ \bibinfo {pages} {245401} (\bibinfo {year} {2015})}\BibitemShut {NoStop}%
\bibitem [{\citenamefont {Zhu}\ \emph {et~al.}(2016)\citenamefont {Zhu}, \citenamefont {Liu}, \citenamefont {Haldane},\ and\ \citenamefont {Sheng}}]{Zhu_2016}%
  \BibitemOpen
  \bibfield  {author} {\bibinfo {author} {\bibfnamefont {W.}~\bibnamefont {Zhu}}, \bibinfo {author} {\bibfnamefont {Z.}~\bibnamefont {Liu}}, \bibinfo {author} {\bibfnamefont {F.~D.~M.}\ \bibnamefont {Haldane}},\ and\ \bibinfo {author} {\bibfnamefont {D.~N.}\ \bibnamefont {Sheng}},\ }\bibfield  {title} {\bibinfo {title} {Fractional quantum hall bilayers at half filling: Tunneling-driven non-abelian phase},\ }\href {https://doi.org/10.1103/PhysRevB.94.245147} {\bibfield  {journal} {\bibinfo  {journal} {Phys. Rev. B}\ }\textbf {\bibinfo {volume} {94}},\ \bibinfo {pages} {245147} (\bibinfo {year} {2016})}\BibitemShut {NoStop}%
\bibitem [{\citenamefont {Parry}(1975)}]{parryElectrostaticPotentialSurface1975a}%
  \BibitemOpen
  \bibfield  {author} {\bibinfo {author} {\bibfnamefont {D.~E.}\ \bibnamefont {Parry}},\ }\bibfield  {title} {\bibinfo {title} {The electrostatic potential in the surface region of an ionic crystal},\ }\bibfield  {journal} {\bibinfo  {journal} {Surface Science}\ }\href {https://doi.org/10.1016/0039-6028(75)90362-3} {10.1016/0039-6028(75)90362-3} (\bibinfo {year} {1975})\BibitemShut {NoStop}%
\bibitem [{\citenamefont {Grzybowski}\ \emph {et~al.}(2000)\citenamefont {Grzybowski}, \citenamefont {Gw{\'o}{\'z}d{\'z}},\ and\ \citenamefont {Br{\'o}dka}}]{grzybowskiEwaldSummationElectrostatic2000a}%
  \BibitemOpen
  \bibfield  {author} {\bibinfo {author} {\bibfnamefont {A.}~\bibnamefont {Grzybowski}}, \bibinfo {author} {\bibfnamefont {E.}~\bibnamefont {Gw{\'o}{\'z}d{\'z}}},\ and\ \bibinfo {author} {\bibfnamefont {A.}~\bibnamefont {Br{\'o}dka}},\ }\bibfield  {title} {\bibinfo {title} {Ewald summation of electrostatic interactions in molecular dynamics of a three-dimensional system with periodicity in two directions},\ }\bibfield  {journal} {\bibinfo  {journal} {Physical Review B}\ }\href {https://doi.org/10.1103/PhysRevB.61.6706} {10.1103/PhysRevB.61.6706} (\bibinfo {year} {2000})\BibitemShut {NoStop}%
\bibitem [{\citenamefont {Needs}\ \emph {et~al.}(2019)\citenamefont {Needs}, \citenamefont {Towler}, \citenamefont {Drummond},\ and\ \citenamefont {López~Ríos}}]{CASINO2019manual}%
  \BibitemOpen
  \bibfield  {author} {\bibinfo {author} {\bibfnamefont {R.~J.}\ \bibnamefont {Needs}}, \bibinfo {author} {\bibfnamefont {M.~D.}\ \bibnamefont {Towler}}, \bibinfo {author} {\bibfnamefont {N.~D.}\ \bibnamefont {Drummond}},\ and\ \bibinfo {author} {\bibfnamefont {P.}~\bibnamefont {López~Ríos}},\ }\href {https://casinoqmc.net/casino_manual_dir/casino_manual.pdf} {\emph {\bibinfo {title} {CASINO User Manual}}} (\bibinfo {year} {2019}),\ \bibinfo {note} {page 186}\BibitemShut {NoStop}%
\bibitem [{\citenamefont {Stern}\ and\ \citenamefont {Howard}(1967)}]{Stern_1967}%
  \BibitemOpen
  \bibfield  {author} {\bibinfo {author} {\bibfnamefont {F.}~\bibnamefont {Stern}}\ and\ \bibinfo {author} {\bibfnamefont {W.~E.}\ \bibnamefont {Howard}},\ }\bibfield  {title} {\bibinfo {title} {Properties of semiconductor surface inversion layers in the electric quantum limit},\ }\href {https://doi.org/10.1103/PhysRev.163.816} {\bibfield  {journal} {\bibinfo  {journal} {Phys. Rev.}\ }\textbf {\bibinfo {volume} {163}},\ \bibinfo {pages} {816} (\bibinfo {year} {1967})}\BibitemShut {NoStop}%
\end{thebibliography}%

\onecolumngrid
\newpage
\makeatletter

\begin{center}
\textbf{\large Supplementary materials for:\\ ``{Electron quasicrystal in semiconductor quantum wells
} ''}
\\[10pt]
Filippo Gaggioli$^{1}$, Pierre-Antoine Graham$^{1}$, Liang Fu$^{1}$\\
\textit{$^1$Department of Physics, Massachusetts Institute of Technology, Cambridge, MA-02139, USA}
\end{center}
\vspace{20pt}



\setcounter{figure}{0}
\setcounter{section}{0}
\setcounter{equation}{0}

\setcounter{secnumdepth}{3}

\renewcommand{\thefigure}{S\arabic{figure}}
\renewcommand{\thesection}{S\arabic{section}}
\renewcommand{\theequation}{S\arabic{equation}}

\section{Model}
We simulate the quantum well in three-dimensional space, with open boundary conditions along $z$ and periodic boundary conditions in the $xy$ plane, where the supercell is spanned by the basis vectors $\mathbf{a}_1, \mathbf{a}_2$, with reciprocal vectors $\mathbf{b}_1, \mathbf{b}_2$ defined by $\mathbf{a}_m \cdot \mathbf{b}_n = 2\pi\delta_{mn}$. For a square and a triangular supercell of linear size $L$, they read
\begin{align}
    \text{square:} \quad
    &\mathbf{a}_1 = L\,(1, 0), \quad \mathbf{a}_2 = L\,(0, 1), \quad
    \mathbf{b}_1 = \frac{2\pi}{L}(1, 0), \quad \mathbf{b}_2 = \frac{2\pi}{L}(0, 1), \nonumber \\
    \text{triangular:} \quad
    &\mathbf{a}_1 = L\,(1, 0), \quad \mathbf{a}_2 = L\,\Big(-\frac{1}{2}, \frac{\sqrt{3}}{2}\Big), \quad
    \mathbf{b}_1 = \frac{2\pi}{L}\Big(1, \frac{1}{\sqrt{3}}\Big), \quad \mathbf{b}_2 = \frac{2\pi}{L}\Big(0, \frac{2}{\sqrt{3}}\Big),\label{eq:supercell_lattice_and_reciprocal}
\end{align}
with supercell areas $A = L^2$ and $A = \sqrt{3}L^2/2$, respectively. In both cases, $L$ is fixed by the density through $A = N\pi r_{\rm 2D}^2$. The electrons are confined to a well of thickness $d_{\rm well}$, extending from $z = -d_{\rm well}/2$ to $z = d_{\rm well}/2$. Global charge neutrality is enforced by two uniformly charged background planes, which we place at the edges of the well, $z = \pm d_{\rm well}/2$, which represent the contribution of donor atoms in realistic quantum wells. For $N$ electrons at positions $\mathbf{r}_i = \boldsymbol{\rho}_i + z_i \hat{\mathbf{z}}$, the Hamiltonian consists of the kinetic energy, the well confinement, and the Coulomb interactions,
\begin{align}
    H = \frac{\hbar^2}{m} \left( -\frac{1}{2}\sum_{i=1}^{N} \nabla^2_i + H_{\rm well} + H_{\rm e\text{-}e} + H_{\rm bg\text{-}e} + H_{\rm bg\text{-}bg} \right), \label{eq:H_SM}
\end{align}
where $m$ is the effective electron mass. Here and in the main text, energies are reported in effective Hartree units, $\text{Ha} = \hbar^2/m\aB^2$.
The individual terms are as follows:

\begin{itemize}

\item[\textbullet] The confinement of the electrons along $z$ is implemented through the smooth step potential
\begin{align}
    H_{\rm well}
    = \sum_{i=1}^{N} \frac{V_{\rm well}}{2}
    \left[
    1 + \tanh\!\left(
    \frac{|z_i| - d_{\rm well}/2}{w}
    \right)
    \right],
     \label{eq:H_detail}
\end{align}
characterized by the well depth $V_{\rm well}$ and the step width $w$. In our simulations, we set $w = 0.05\, d_{\rm well}$ and choose $V_{\rm well}$ to be the largest energy scale in the system.

\item[\textbullet] The electron--electron interaction is
\begin{align}
    H_{\rm e\text{-}e} = \frac{N\phi_0}{2a_B} + \frac{1}{2a_B}\sum_{i=1}^{N} \sum_{j \neq i} \phi(\mathbf{r}_i - \mathbf{r}_j),\label{eq:H_ee}
\end{align}
where the Coulomb kernel $\phi(\mathbf{r})$ and the Madelung constant $\phi_0$ for a geometry that is periodic in the plane and open along $z$ are given by the Ewald summation formula for a slab~\cite{parryElectrostaticPotentialSurface1975a, grzybowskiEwaldSummationElectrostatic2000a},
\begin{align}
    \phi(\mathbf{r}) &{}= \sum_{\mathbf{L}} \frac{\text{erfc}[\alpha\sqrt{(\boldsymbol{\rho} + \mathbf{L})^2 + z^2}]}{\sqrt{(\boldsymbol{\rho} + \mathbf{L})^2 + z^2}} - \frac{2\sqrt{\pi}}{A} \left(\frac{1}{\alpha} e^{-\alpha^2 z^2} + \sqrt{\pi}z\text{erf}(\alpha z)\right)\nonumber\\
    &{}+\frac{\pi}{A} \sum_{\mathbf{G}\neq \mathbf{0}} \frac{e^{i \mathbf{G}\cdot \boldsymbol{\rho}}}{|\mathbf{G}|} \left( e^{|\mathbf{G}|z} \text{erfc}\left(\frac{|\mathbf{G}|}{2\alpha} +\alpha z\right) + e^{-|\mathbf{G}|z} \text{erfc}\left(\frac{|\mathbf{G}|}{2\alpha} - \alpha z\right)  \right),\nonumber\\[0.3cm]
    \phi_0 &{}= - \frac{2\sqrt{\pi}}{A} \frac{1}{\alpha} - \frac{2\alpha}{\sqrt{\pi}} + \sum_{\mathbf{L}\neq \mathbf{0}} \frac{\text{erfc}[\alpha|\mathbf{L}|]}{|\mathbf{L}|} + \frac{2\pi}{A}\sum_{\mathbf{G}\neq \mathbf{0}} \frac{1}{|\mathbf{G}|} \text{erfc}\left(\frac{|\mathbf{G}|}{2\alpha}\right).\label{eq:ewald}
\end{align}
Here, $A$ is the supercell area, $\mathbf{L} \in \{n\mathbf{a}_1 + m\mathbf{a}_2 : (n, m) \in \mathbb{Z}^2\}$ runs over the supercell lattice, and $\mathbf{G} \in \{n \mathbf{b}_1 + m \mathbf{b}_2 : (n, m)\in \mathbb{Z}^2\}$ over the associated reciprocal lattice. We use square and triangular supercells and set $\alpha = 2.4/L$ for efficient convergence~\cite{CASINO2019manual}.

\item[\textbullet] Owing to the periodic boundary conditions, each background plane acts on all charges as an infinite, uniformly charged sheet. The background--electron and background--background interactions are therefore
\begin{align}
    H_{\rm bg\text{-}e} = \frac{N\pi}{A a_B} \sum_{i=1}^{N} \left(|z_i - d_{\rm well}/2| + |z_i + d_{\rm well}/2|\right), \qquad H_{\rm bg\text{-}bg} = - \frac{N^2 \pi d_{\rm well}}{2 A a_B}. \label{eq:H_bg}
\end{align}

\end{itemize}

\section{Neural Network}
To represent the wave function $\Psi_{\theta}(\mathbf{r}_1, \dots, \mathbf{r}_N)$, we use the attention-based architecture of Ref.~\cite{Geier_2025}, itself adapted from the Psiformer~\cite{vonGlehn_2023}, to which we refer for details. Below we describe the modifications introduced in this work.

Writing $\mathbf{r}_i = (\boldsymbol{\rho}_i, z_i)$, periodicity in the plane is built in through the feature layer~\cite{Cassella_2023}
\begin{align}
    \mathbf{f}(\mathbf{r}_i) = \left(
        \sin(\mathbf{b}_1 \cdot \boldsymbol{\rho}_i), \
        \sin(\mathbf{b}_2 \cdot \boldsymbol{\rho}_i), \
        \cos(\mathbf{b}_1 \cdot \boldsymbol{\rho}_i), \
        \cos(\mathbf{b}_2 \cdot \boldsymbol{\rho}_i), \
        z_i/d_{\rm well}
    \right),
\end{align}
where $\mathbf{b}_{1,2}$ are the reciprocal vectors of the supercell $\mathbf{a}_{1,2}$ defined in Eqs.~\eqref{eq:supercell_lattice_and_reciprocal}, and $d_{\rm well}$ is the well thickness.

\paragraph{Spatial attention bias.}
In each attention layer $\ell$ and head $h$, we add to the attention logits a learnable bias proportional to the (periodic) in-plane electron--electron distance \cite{Gaggioli_2026},
\begin{align}
    A_{ij}^{\ell h} = \underset{j}{\mathrm{softmax}} \left(
        \frac{\mathbf{q}_i^{\ell h} \cdot \mathbf{k}_j^{\ell h}}{\sqrt{d_h}}
        - \gamma_{\ell h}\, d(\boldsymbol{\rho}_i - \boldsymbol{\rho}_j)
    \right),
     \label{eq:spatial_attention}
\end{align}
where $\mathbf{q}_i^{\ell h}$, $\mathbf{k}_j^{\ell h}$ are the queries and keys of head dimension $d_h$, and $\gamma_{\ell h}$ is a per-head parameter, initialized to $\sqrt{2}/r_0$ with $r_0 = \sqrt{A/N}$ the mean interparticle distance and $A$ the supercell area. Heads with large $\gamma_{\ell h}$ thus attend more markedly to nearby electrons, while $\gamma_{\ell h} \to 0$ recovers standard attention.
To respect the periodic boundary conditions, we use the norm of Ref.~\cite{Cassella_2023}, defined as
\begin{align}
    d(\boldsymbol{\rho}) = \frac{1}{2\pi}\, \Big\lVert \sum_{m=1}^{2} \mathbf{a}_m \left( e^{i \mathbf{b}_m \cdot \boldsymbol{\rho}} - 1 \right) \Big\rVert ,
     \label{eq:periodic_distance}
\end{align}
which is smooth and differentiable, periodic on the supercell, and reduces to $|\boldsymbol{\rho}|$ at short distances. We remark that only the in-plane component of the interparticle distance is fed into the spatial attention~\eqref{eq:spatial_attention}, in order to preserve inter-layer attention.

\paragraph{Wave function.}
The output of the last layer is linearly mapped to complex generalized orbitals $\phi_j^{k}(\mathbf{r}_i; \{\mathbf{r}_{\neq i}\})$, $1 \leq j \leq N$, $1 \leq k \leq N_{\rm det}$. Including a Gaussian envelope that bounds the electrons inside the well, the wave function reads
\begin{align}
    \Psi_{\theta}(\mathbf{r}_1, \dots, \mathbf{r}_N) =
    e^{i \boldsymbol{\theta} \cdot \sum_{i} \boldsymbol{\rho}_i}\,
    \exp\Big( -\sigma_z \sum_{i=1}^{N} z_i^2 \Big)
    \sum_{k=1}^{N_{\rm det}} \det_{ij}\big[ \phi_j^{k}(\mathbf{r}_i; \{\mathbf{r}_{\neq i}\}) \big],
     \label{eq:ansatz}
\end{align}
with trainable $\sigma_z$, initialized to $\pi^2/d_{\rm well}^2$. The global phase implements twisted boundary conditions with twist $\boldsymbol{\theta}$; it leaves $|\Psi_\theta|^2$ unchanged and enters only through the kinetic energy. Unless stated otherwise, $\boldsymbol{\theta} = 0$. For the spin-polarized simulations presented in this work, no Jastrow factor is used.

\paragraph{Sampling.}
The expectation values entering the optimization are estimated over an ensemble of walkers (Table~\ref{tab:train-hparams}), each evolved by an independent Metropolis--Hastings chain targeting $|\Psi_\theta|^2$. We use all-electron moves: at each step, every coordinate of every electron is displaced simultaneously by an isotropic Gaussian increment of width $\delta$, $\mathbf{r}_i \to \mathbf{r}_i + \delta\,\boldsymbol{\eta}_i$ with $\boldsymbol{\eta}_i \sim \mathcal{N}(0, \mathbb{1}_3)$, after which the in-plane coordinates are mapped back into the supercell. The proposal is symmetric, so the move is accepted with probability $\min\!\big(1, |\Psi_\theta(\mathbf{R}')|^2/|\Psi_\theta(\mathbf{R})|^2\big)$. 
Between consecutive parameter updates, each chain performs a number of ``thinning'' steps ($500$ for the $N = 78$ runs) to decorrelate the samples. The width $\delta$ is adapted during training to keep the acceptance rate between $50\%$ and $55\%$.
Before the start of training, the chains are equilibrated by a burn-in phase without parameter updates.

The network is optimized with KFAC~\cite{vonGlehn_2023}; the hyperparameters used to produce the phase diagram in Fig.~\ref{fig:phase_diagrams_combined} are listed in Table~\ref{tab:train-hparams}.

\begin{table}[h!]
  \centering
  \caption{Architecture and training hyperparameters.}
  \label{tab:train-hparams}%
  \begin{tabular}{@{}l@{\hspace{1.5em}}l@{\hspace{3em}}l@{\hspace{1.5em}}l@{}}
    \toprule
    \multicolumn{2}{@{}l}{Architecture} & \multicolumn{2}{l}{Optimization and sampling} \\
    \midrule
    Number of electrons $N$ & $78$ & Optimizer & KFAC \\
    Number of layers $L$ & $5$ & Learning rate & $0.1$--$1.0$ \\
    Attention heads per layer & $\leq 8$ & KFAC damping & $1\times10^{-4}$ \\
    Head dimension $d_h$ & $\leq 32$ & KFAC norm constraint & $1\times10^{-3}$ \\
    Embedding width & $\leq 256$ & Local-energy clipping & $5$ \\
    MLP hidden width & $\leq 256$ & Batch size (walkers) & $2048$--$4096$ \\
    Activation & GELU & Training iterations & $5\times10^{4}$--$1.5\times10^{5}$ \\
    Layer norm & True & MCMC steps between iterations & $500$ \\
    Number of determinants $N_{\rm det}$ & $2$ & Precision & FP$32$ \\
    Initial $\gamma_{\ell h}$ & $\sqrt{2}/r_0$ & Number of parameters & $<1$M \\
    \bottomrule
  \end{tabular}
\end{table}

\section{One-body reduced density matrix}

In this section, we give the expression used to evaluate the one-body reduced density matrix (1RDM) elements displayed in Fig. \ref{fig:mono_to_bilayer}. We work in a single particle basis that is the product of plane-waves in the $x, y$ plane, and eigenstates of an infinite quantum well of width $d_{\rm well}$ along the $z$ direction,
\begin{align}
    \Phi_{n_z, \kb} = \sqrt{\frac{2}{d_{\rm well}A}}e^{i \kb \cdot \boldsymbol{\rho}} \sin \left(\frac{n_z\pi}{d_{\rm well}}z - \frac{n_z\pi}{2}\right) \label{eq:infinite_well_basis}
\end{align}
where $n_z \geq 1$ and $\mathbf{k}$ is an in-plane reciprocal lattice vector. In this basis, the 1RDM has elements
\begin{align}
    \rho_1(n_z, \kb; n_z', \kb') &= \sum_{i=1}^{N} \int \text{d}^3\mathbf{r}_{i} \text{d}^3\mathbf{r}'_i \left( \prod_{j \neq i } \text{d}^3\mathbf{r}_j \right)\ \Psi(\cdots,\ \mathbf{r}_i,\ \cdots) \Psi^{*}(\cdots,\  \mathbf{r}_i',\ \cdots) \Phi^{*}_{n_z, \kb}(\mathbf{r}_i) \Phi_{n_z', \kb'}(\mathbf{r}_i')\nonumber\\
    &= \sum_{i=1}^{N} \int \text{d}^3\mathbf{r}_i \left( \prod_{j \neq i } \text{d}^3\mathbf{r}_j \right) \  |\Psi(\cdots,\ \mathbf{r}_i,\ \cdots)|^2 \int \text{d}'\mathbf{r}_i  \frac{\Psi^{*}(\cdots,\ \mathbf{r}_i',\ \cdots)}{\Psi^{*}(\cdots,\ \mathbf{r}_i,\ \cdots)} \Phi^{*}_{n_z, \kb}(\mathbf{r}_i) \Phi_{n_z', \kb'}(\mathbf{r}_i').
\end{align}
On the second line, the elements are expressed in a form that can be evaluated by Monte Carlo sampling a batch of configurations $\mathcal{B}$  from the probability distribution $|\Psi|^2$, yielding
\begin{align}
     \rho_1(n_z, \mathbf{k}; n_z', \mathbf{k}') &\approx \sum_{i=1}^N\sum_{\{\mathbf{r}_i\} \in \mathcal{B}} \int \frac{\text{d}'\mathbf{r}_i }{N_\mathcal{B}} \frac{\Psi^{*}(\cdots,\ \mathbf{r}_i',\ \cdots)}{\Psi^{*}(\cdots,\ \mathbf{r}_i,\ \cdots)} \Phi^{*}_{n_z, \mathbf{k}}(\mathbf{r}_i) \Phi_{n_z', \mathbf{k}'}(\mathbf{r}_i'),
\end{align}
where $N_\mathcal{B}$ is the batch size. In particular, the subband-resolved momentum space occupations $n(\mathbf{k})$ displayed in Fig.~\ref{fig:mono_to_bilayer}~$(c)$ an $(d)$ in the main text are given by the diagonal elements
\begin{align}
     n(\mathbf{k};n_z) &\approx \sum_{i=1}^N\sum_{\{\mathbf{r}_i\} \in \mathcal{B}} \int \frac{\text{d}'\mathbf{r}_i}{N_\mathcal{B}} \, \frac{\Psi^{*}(\cdots,\ \mathbf{r}_i',\ \cdots)}{\Psi^{*}(\cdots,\ \mathbf{r}_i,\ \cdots)} \Phi^{*}_{n_z, \mathbf{k}}(\mathbf{r}_i) \Phi_{n_z, \mathbf{k}}(\mathbf{r}_i'). 
\end{align}


\section{Energetics of the quasicrystal stacking}
\label{sec:energetics}
In this section, we detail the energy comparison that allows us to establish the ground state nature of the quasicrystal stacking for sufficiently large quantum well thickness. 
To check that our conclusions are robust against strain and system
size, we compare the two stackings in both supercells of
Eq.~\eqref{eq:supercell_lattice_and_reciprocal}, each of which accommodates the
honeycomb and quasicrystal candidates on an equal footing so as not to bias the
comparison. The square supercell with $N=30$ electrons hosts, beyond the (strained)
honeycomb stacking, a quasicrystal made of two identical (strained) triangular lattices of $15$ electrons
each, rotated by exactly $30^\circ$. The triangular supercell
with $N=78$ electrons lets the $39$ electrons of each layer form an unstrained
triangular lattice with two possible orientations, with primitive vectors
$\mathbf t_{1,2}$ and $\mathbf t'_{1,2}$ related to the supercell vectors by
$\mathbf a_1=5\mathbf t_1+2\mathbf t_2$ and $\mathbf a_1=2\mathbf t'_1+5\mathbf t'_2$.
Stacking two such rotated lattices realizes a quasicrystal \textit{approximant} with
twist angle $\theta=\arccos(23/26)\approx 27.8^\circ$~\cite{Pleasants_1996,Mele_2010}
(the next approximant requires $194$ electrons and has
$\theta=\arccos(169/194)\approx 29.4^\circ$). In the same supercell, the honeycomb stacking
consists of two identical triangular layers, the top one shifted by
$(\mathbf t_1+\mathbf t_2)/3$ with respect to the bottom one.

\begin{figure*}[t]
    \centering
    \includegraphics[width=\textwidth]{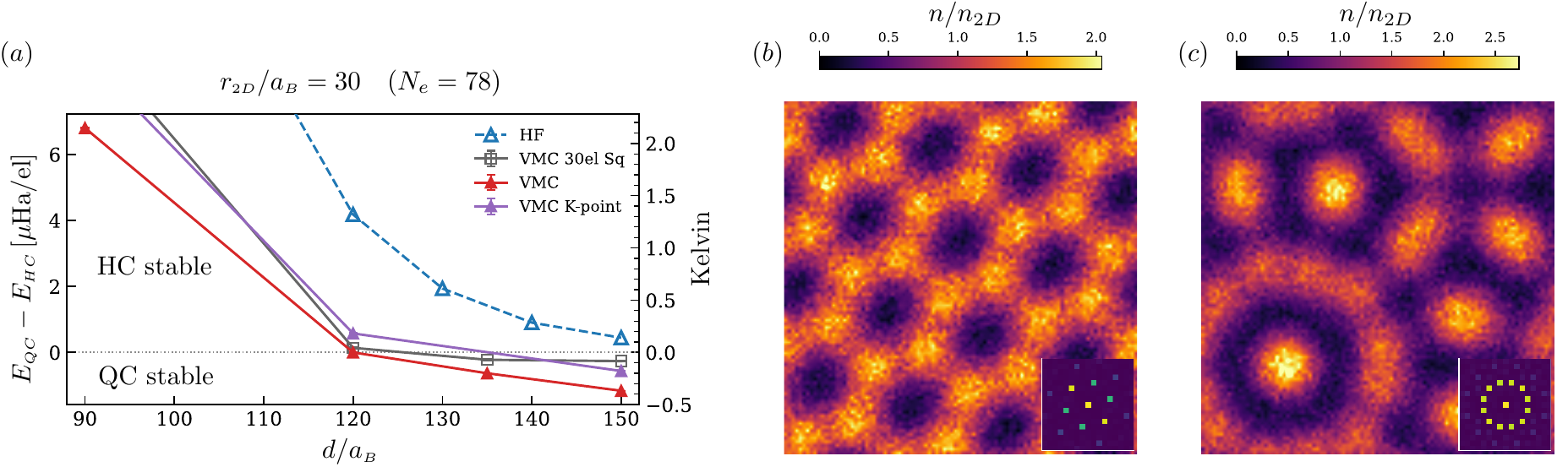}
    \caption{\textbf{Systematic energy comparison, including square supercell and twisted boundary conditions:} (a) NN-VMC energy difference (solid lines) between the honeycomb (HC) and quasicrystal (QC) stackings, for triangular (red and purple) and square (grey) supercells with $N=78$ and $N=30$ particles, respectively, at fixed $r_{2D}/\aB=30$.
    The red and purple line show the energy differences for periodic and twisted ($K$-point) boundary conditions, respectively. The Hartree-Fock results for the triangular supercell with $N=78$ electrons are shown with dashed lines.
    (b) and (c) show representative NN-VMC real-space charge distribution corresponding to the honeycomb and quasicrystal stackings in a square supercell with $N=30$ particles. }
    \label{fig:sq_vs_hc_vs_qc_supplementary}
\end{figure*}

Upon random initialization, the NN-VMC training can end up in either the honeycomb or quasicrystal stacking. Direct comparison of the converged energies can then establish the nature of the ground state, as shown in Fig.~\ref{fig:sq_vs_hc_vs_qc_supplementary}.
The corresponding absolute values, for the considered combinations of system sizes, supercell geometries and boundary conditions (with the $K$-point twist obtained by setting $\boldsymbol{\theta} = (\mathbf{b}_1+\mathbf{b}_2)/3$ in the neural network ansatz Eq.~\eqref{eq:ansatz}), are reported in Table \ref{tab:qc-hc-energies} along with the Hartree-Fock results (see Sec.~\ref{sec:hf} below). The honeycomb and quasicrystal charge density distributions for the square supercell are shown in Fig.~\ref{fig:sq_vs_hc_vs_qc_supplementary} $(b)$ and $(c)$ in this supplementary.

\begin{table*}[h!]
  \centering
  \caption{Absolute energies of the HC and QC phases at $r_{\scriptscriptstyle 2D}/\aB = 30$, in $\mu$Ha/electron, and their difference $\Delta E = E_{\rm QC} - E_{\rm HC}$ as shown in Figs.~\ref{fig:sq_vs_hc_vs_qc} and \ref{fig:sq_vs_hc_vs_qc_supplementary}. VMC energies and statistical errors obtained by averaging over $500$ iterations after convergence.
  }
  \label{tab:qc-hc-energies}
  \begin{tabular*}{\textwidth}{@{\extracolsep{\fill}}lrrrrlrrrr@{}}
    \toprule
    Method & $t_s$ & $E_{\rm HC}$ & $E_{\rm QC}$ & $\Delta E$ & Method & $t_s$ & $E_{\rm HC}$ & $E_{\rm QC}$ & $\Delta E$ \\
    \midrule
    VMC, $N=78$ & $90$ & $1011.01(6)$ & $1017.81(6)$ & $6.8(0)$ & HF, $N=78$ & $80$ & $508.894$ & $582.242$ & $73.35$ \\
    Triangular supercell & $120$ & $3894.80(1)$ & $3894.80(1)$ & $0.00(0)$ & Triangular supercell & $90$ & $1496.477$ & $1534.374$ & $37.90$ \\
     & $135$ & $5342.78(1)$ & $5342.15(0)$ & $-0.63(2)$ & & $100$ & $2472.350$ & $2491.012$ & $18.66$ \\
     & $150$ & $6793.95(1)$ & $6792.78(9)$ & $-1.16(2)$ & & $110$ & $3444.026$ & $3452.943$ & $8.92$ \\
    \cmidrule(r){1-5}
    VMC ($K$-point), $N=78$ & $90$ & $1015.04(6)$ & $1024.02(9)$ & $8.9(8)$ & & $120$ & $4414.378$ & $4418.556$ & $4.18$ \\
    Triangular supercell & $120$ & $3899.73(8)$ & $3900.31(4)$ & $0.57(6)$ & & $130$ & $5385.046$ & $5386.982$ & $1.94$ \\
     & $150$ & $6798.97(3)$ & $6798.41(0)$ & $-0.56(2)$ & & $140$ & $6357.000$ & $6357.903$ & $0.90$ \\
    \cmidrule(r){1-5}
    VMC, $N=30$ & $90$ & $982.2(3)$ & $991.8(7)$ & $9.6(4)$ & & $150$ & $7330.581$ & $7331.025$ & $0.44$ \\
    Square supercell & $120$ & $3859.6(9)$ & $3859.8(3)$ & $0.1(4)$ & & & & & \\
     & $135$ & $5308.7(6)$ & $5308.5(3)$ & $-0.2(2)$ & & & & & \\
     & $150$ & $6754.31(8)$ & $6754.05(5)$ & $-0.26(3)$ & & & & & \\
    \bottomrule
  \end{tabular*}
\end{table*}

\section{Hartree--Fock analysis} \label{sec:hf}
As a correlation-free reference for the NN-VMC results, we treat the
Hamiltonian~\eqref{eq:H_SM} in the unrestricted Hartree--Fock (HF) approximation. We
consider spin-polarized electrons in the two supercells of
Eq.~\eqref{eq:supercell_lattice_and_reciprocal}, with $N=30$ electrons in the square
supercell used for the NN-VMC results of Fig.~\ref{fig:mono_to_bilayer} and for the
classical configurations of Fig.~\ref{fig:sq_vs_hc_vs_qc}(b,c), and $N=78$ electrons
in the triangular supercell used for Fig.~\ref{fig:bilayer_WC}, which hosts 39 sites
per layer.

\begin{figure}[t]
  \centering
  \includegraphics[width=0.8\textwidth]{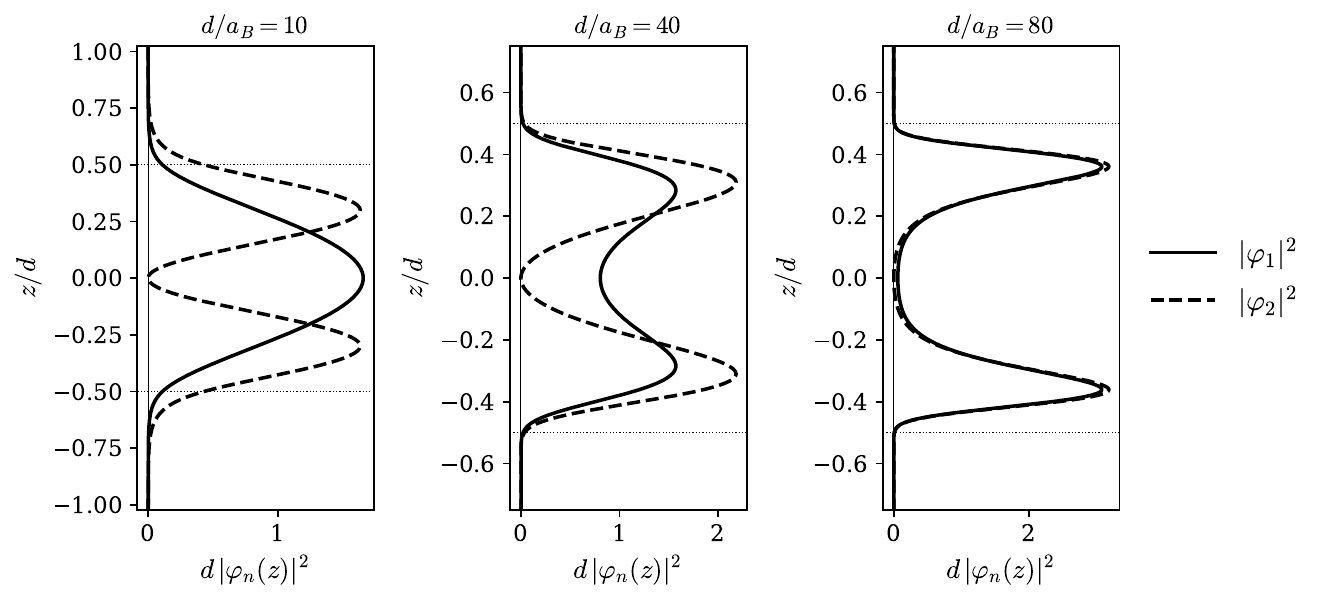}
  \caption{\textbf{Hartree-Fock subbands:} Density profiles $|\varphi_n(z)|^2$ of the two lowest subband functions of
  the HF basis~\eqref{eq:HF_basis}, in scale-free units, for three well widths $d$
  ($N=30$, $r_{2D}/a_B=20$). Dotted lines mark the well edges $z=\pm d/2$.}
  \label{fig:hf_subbands}
\end{figure}

\textit{Basis.} The single-particle orbitals are expanded in products of plane waves and
subband functions,
\begin{equation}
  \Phi_{n,\mathbf G}(\mathbf r)=\frac{1}{\sqrt A}\,e^{i\mathbf G\cdot\bm\rho}\,
  \varphi_{n}(z),\qquad |\mathbf G|\le G_c,\quad n=1,\dots,n_z^{\max},
  \label{eq:HF_basis}
\end{equation}
where $\mathbf G\in\{n\mathbf b_1+m\mathbf b_2:(n,m)\in\mathbb Z^2\}$ runs over the
reciprocal lattice of the supercell up to a cutoff $G_c$. In contrast
to the infinite-well states of Eq.~\eqref{eq:infinite_well_basis}, the subband
functions $\varphi_{n}$ are adapted to the interacting problem. They are chosen as
the lowest $n_z^{\max}=6$ eigenfunctions of the one-dimensional Hartree Hamiltonian
\begin{equation}
  h_z=\frac{\hbar^2}{m}\left(-\frac12\partial_z^2+V_{\rm well}(z)+V_{\rm bg}(z)
  -\frac{2\pi}{a_B}\int dz'\,n_0(z')\,|z-z'|\right),
  \label{eq:HF_hz}
\end{equation}
where $V_{\rm well}$ and $V_{\rm bg}$ denote the one-body potentials entering
$H_{\rm well}$ and $H_{\text{bg-e}}$, and $n_0(z)$ is the self-consistent areal density of
$N$ electrons occupying the lowest subband, as in the Stern--Howard treatment of
inversion layers~\cite{Stern_1967}. As shown in Fig.~\ref{fig:hf_subbands}, the lowest
subband already reproduces the flat-topped charge profile induced by the Hartree
repulsion, and the higher ones provide the antisymmetric and breathing deformations
needed to describe a bilayer, so that a few functions suffice.

Being eigenfunctions of $h_z$, the subband functions are orthonormal,
$\int dz\,\varphi_n(z)\varphi_m(z)=\delta_{nm}$, and so are the basis
functions~\eqref{eq:HF_basis}. In this basis, the $N$ occupied orbitals of the Slater
determinant read
$\psi_j(\mathbf r)=\sum_{\mathbf G,n}C_{\mathbf G n,j}\,\Phi_{n,\mathbf G}(\mathbf r)$,
$j=1,\dots,N$, where here and below $n,m,n',m'\in\{1,\dots,n_z^{\max}\}$ label subbands.
We do not require these orbitals to be orthonormal. The Slater determinant depends only
on the subspace they span, and is fully characterized by the
one-body density matrix
\begin{equation}
P=CS^{-1}C^\dagger,\qquad S=C^\dagger C,
  \label{eq:HF_P}
\end{equation}
where $C$ is the $N_Gn_z^{\max}\times N$ matrix of coefficients $C_{\mathbf G n,j}$ and
$S$ the $N\times N$ overlap matrix of the orbitals.

\textit{Energy functional.} In terms of the density matrix~\eqref{eq:HF_P}, the HF
energy is the sum of four terms,
\begin{equation}
  E_{\rm HF}=\frac{\hbar^2}{m}\left(\frac{N\phi_0}{2a_B}+H_{\text{bg-bg}}\right)
  +E_{\rm 1b}+E_{\rm H}+E_{\rm x}.
  \label{eq:HF_energy}
\end{equation}
The one-body energy is diagonal in $\mathbf G$,
\begin{equation}
  E_{\rm 1b}=\frac{\hbar^2}{m}\sum_{\mathbf G}\sum_{nm}
  \left(\frac{|\mathbf G|^2}{2}\,\delta_{nm}+t_{nm}\right)P_{\mathbf G,m;\mathbf G,n},
  \qquad
  t_{nm}=\int dz\,\varphi_n(z)\left[-\frac12\partial_z^2+V_{\rm well}(z)
  +V_{\rm bg}(z)\right]\varphi_m(z),
  \label{eq:HF_h}
\end{equation}
where the first term is the in-plane kinetic energy and $t_{nm}$ contains the
out-of-plane kinetic energy and the confinement and background potentials. The Hartree
and exchange energies are
\begin{align}
  E_{\rm H}&=\frac12\sum_{\mathbf G,\mathbf G',\mathbf q}\sum_{nm,n'm'}
  U_{nm,n'm'}(\mathbf q)\,P_{\mathbf G+\mathbf q,n;\mathbf G,m}\,
  P_{\mathbf G'-\mathbf q,n';\mathbf G',m'},
  \label{eq:HF_hartree}\\
  E_{\rm x}&=-\frac12\sum_{\mathbf G,\mathbf G',\mathbf q}\sum_{nm,n'm'}
  U_{nm,n'm'}(\mathbf q)\,P_{\mathbf G+\mathbf q,n;\mathbf G',m'}\,
  P_{\mathbf G'-\mathbf q,n';\mathbf G,m},
  \label{eq:HF_exchange}
\end{align}
where $\mathbf q$ runs over the reciprocal lattice of the supercell. The two terms
differ only by the exchange of the second indices of the density matrices. The
interaction matrix elements between subband pairs follow from the in-plane Fourier
transform of the Coulomb interaction, $2\pi e^{-q|z-z'|}/q$ with $q=|\mathbf q|$,
\begin{equation}
  U_{nm,n'm'}(\mathbf q)=\frac{\hbar^2}{m a_B}\,\frac{2\pi}{A}\times\begin{cases}
  \dfrac{1}{q}\displaystyle\int dz\,dz'\,\varphi_n(z)\varphi_m(z)\,e^{-q|z-z'|}\,
  \varphi_{n'}(z')\varphi_{m'}(z'), & q\neq0,\\[2ex]
  -\displaystyle\int dz\,dz'\,\varphi_n(z)\varphi_m(z)\,|z-z'|\,
  \varphi_{n'}(z')\varphi_{m'}(z'), & q=0.
  \end{cases}
  \label{eq:HF_U}
\end{equation}
The $q=0$ kernel is the finite remainder of the slab Ewald interaction after cancellation
of its divergence by the background. The constants $N\phi_0/2a_B$ and $H_{\text{bg-bg}}$ of
Eqs.~\eqref{eq:H_ee} and~\eqref{eq:H_bg} do not affect the HF solution, but allow direct comparison with the VMC energy.
\textit{Minimization.} To find the optimal Slater determinant within the
basis~\eqref{eq:HF_basis}, we minimize $E_{\rm HF}$ directly with respect to the
coefficient matrix $C$, through the density matrix~\eqref{eq:HF_P}, with no symmetry
imposed on the orbitals within the supercell. The minimization is performed with the
L-BFGS algorithm acting on $(\operatorname{Re}C,\operatorname{Im}C)$, using the  gradient
\begin{equation}
  \frac{\partial E_{\rm HF}}{\partial C^*}=(1-P)\,F\,C\,S^{-1},
  \label{eq:HF_gradient}
\end{equation}
where $F$ is the Fock matrix, $F_{\mathbf G n;\mathbf G'm}=\partial E_{\rm HF}/\partial
P_{\mathbf G',m;\mathbf G,n}$, obtained by differentiating
Eqs.~\eqref{eq:HF_h}--\eqref{eq:HF_exchange}, and $P$ and $S$ are given by
Eq.~\eqref{eq:HF_P}. The iterations stop when the norm of this gradient falls below
$10^{-6}$, the energy error being quadratic in it.

\textit{Initial states.} The HF phase diagram of Fig.~\ref{fig:phase_diagrams_combined}
is obtained by comparing, at each $(r_{2D}/a_B,\,d/a_B)$, the energies of four
candidates, the triangular monolayer, the staggered square bilayer, the honeycomb
bilayer and the quasicrystal. Each candidate is specified by $N$ in-plane sites
$\bm\rho_j$ with layer labels $l_j\in\{{\rm b},{\rm t}\}$, chosen as so to fit the square and triangular supercell at $N=30$ and $78$ particles, and is initialized with one orbital per site. In plane,
orbital $j$ is a Gaussian of width $\sigma=a_B$ centred on $\bm\rho_j$, made periodic
by summing over the supercell lattice. Out of plane, it is the profile
$\chi_{l_j}(z)=\sum_n u^{(l_j)}_n\varphi_n(z)$ of its layer, where $u^{({\rm b})}$ and
$u^{({\rm t})}$ are the eigenvectors of $z_{nm}=\int dz\,\varphi_n\,z\,\varphi_m$
restricted to $\{\varphi_1,\varphi_2\}$, so that
$\chi_{\rm b,t}\approx(\varphi_1\mp\varphi_2)/\sqrt2$ lie in the lower and upper halves
of the well. For the monolayer, $u_n=\delta_{n1}$. The corresponding coefficients are
\begin{equation}
  C^{(0)}_{\mathbf G n,j}=e^{-\sigma^2|\mathbf G|^2/4}\,e^{-i\mathbf G\cdot\bm\rho_j}\,u^{(l_j)}_n.
  \label{eq:HF_seed}
\end{equation}
Since $P$ depends only on the span of the columns of $C$, $C^{(0)}$ needs no
orthonormalization.

\begin{figure}[t]
  \centering
  \includegraphics[width=\textwidth]{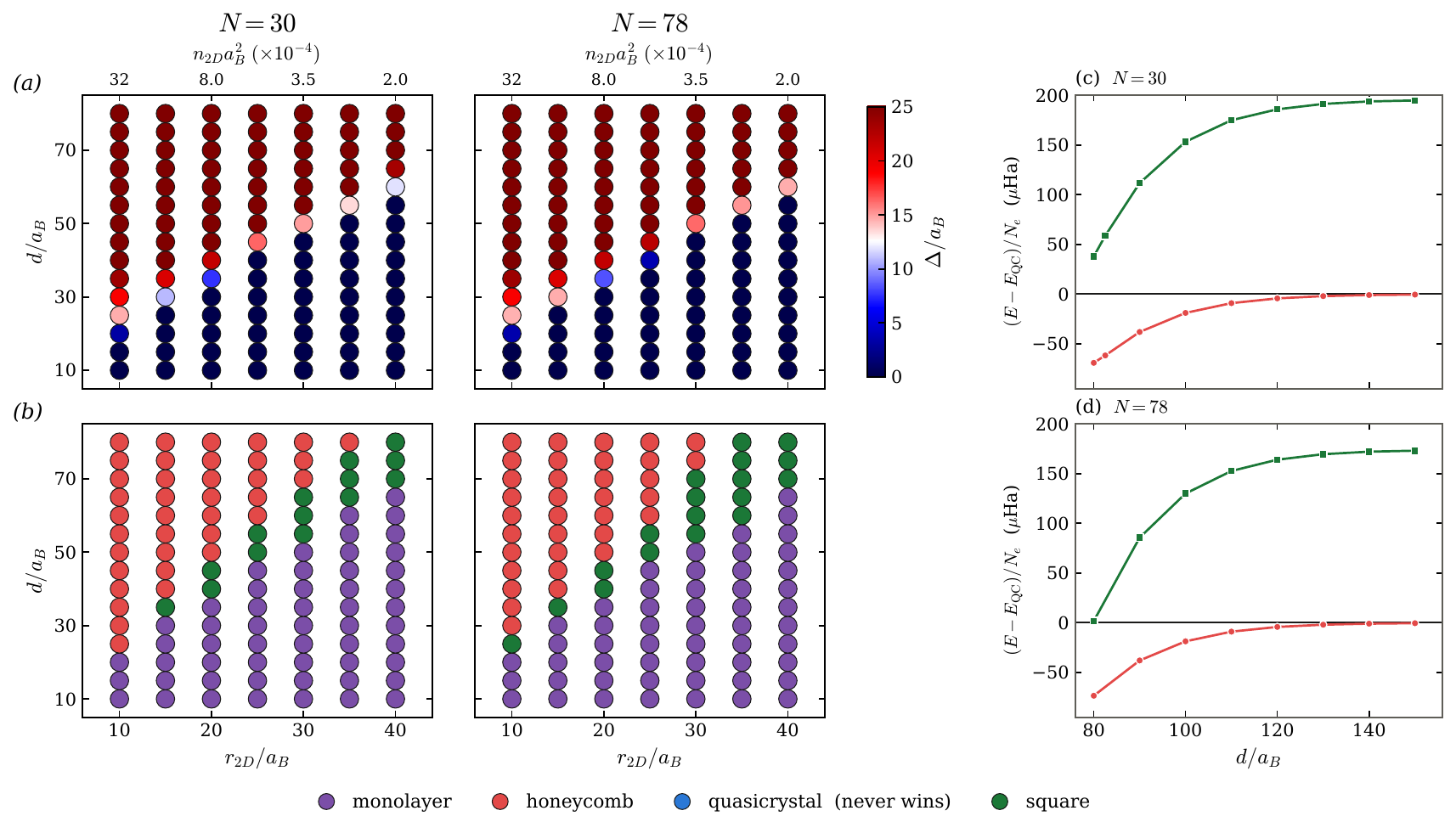}
  \caption{\textbf{Hartree--Fock phase diagram} for $N=30$ (square supercell) and
  $N=78$ (triangular supercell) on the $(r_{2D}/a_B,\,d/a_B)$ grid of Fig.~\ref{fig:phase_diagrams_combined}. (a)~Layer separation $\Delta$ of the HF
  ground state, with the color scale of Fig.~\ref{fig:phase_diagrams_combined}(a).
  $\Delta=0$ corresponds to a monolayer. (b)~Crystal structure of the HF ground state.
  The quasicrystal is never the lowest-energy state. (c,\,d)~Energy per electron of
  the honeycomb (red) and square (green) bilayers relative to the quasicrystal at
  $r_{2D}/a_B=30$ as a function of $d/a_B$, for the square $(c)$ and triangular $(d)$ supercells. The honeycomb
  stays below the quasicrystal at every $d$, with a splitting that decays
  exponentially with $d/r_{\scriptscriptstyle 2D}$.}
  \label{fig:hf_phase_diagram}
\end{figure}

\textit{Results.} Since the HF solution is unrestricted within the supercell, each
initial state is free to relax, to change its registry, or to evolve into a different
state. We therefore classify each converged state by its final charge density, and
retain the stacking with lowest energy at each point. The resulting HF phase diagram is shown in
Fig.~\ref{fig:hf_phase_diagram} for both supercells, in the same layout as
Fig.~\ref{fig:phase_diagrams_combined}. Panel (a) displays the layer separation
$\Delta$ of the HF ground state and panel (b) its optimal crystal structure. Panels
(c) and (d) follow the honeycomb and square energies relative to the quasicrystal
along the line $r_{2D}/a_B=30$ of the NN-VMC calculations, for the two supercells.

Upon increasing the well thickness $d$ at fixed $r_{2D}$, the HF ground state evolves
from the triangular monolayer to the square bilayer and then to the honeycomb bilayer,
in agreement with the sequence of bilayer structures found by NN-VMC
(Fig.~\ref{fig:bilayer_WC}). In contrast, the quasicrystal is never the HF ground state in either
supercell. Wherever both the honeycomb and the quasicrystal are stable solutions, the
honeycomb has lower HF energy, by an amount that decays exponentially with $d$ along
the $r_{2D}/a_B=30$ line of Fig.~\ref{fig:hf_phase_diagram}(c,d). The quasicrystal
found by NN-VMC in wide quantum wells is therefore not stabilized by exchange or
electrostatic effects captured within a single Slater determinant. This supports its
origin in the zero-point energy of the out-of-phase phonons discussed in the main text, a collective-mode effect beyond the reach of HF.

\section{Classical analysis of the bilayer stacking}\label{sec:classical}
\begin{figure*}[t]
    \centering
    \includegraphics[width=\textwidth]{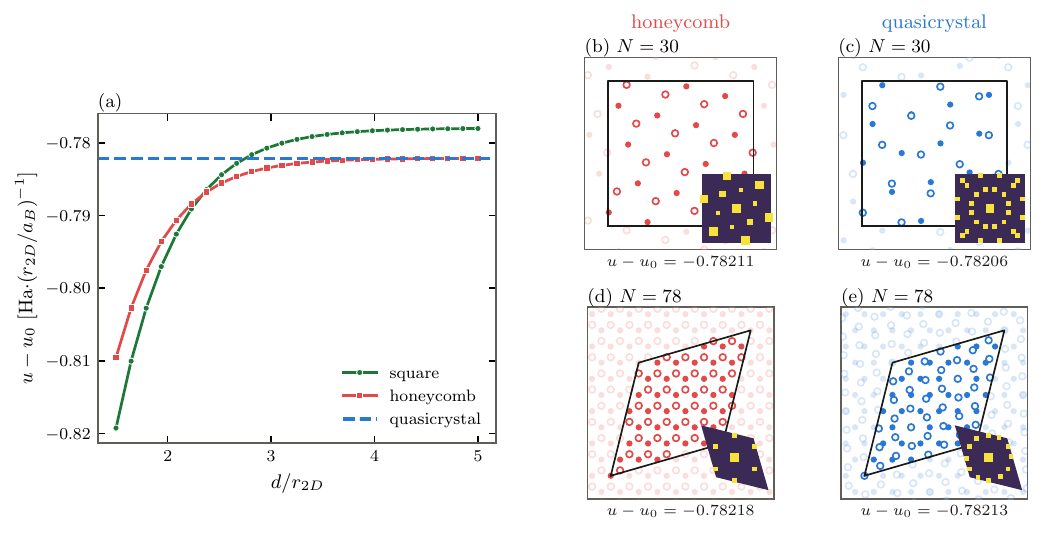}
    \caption{Classical energetics of the bilayer.
    (a) Energy per electron $u$ of the infinite square (green) and honeycomb (red) stackings,
    relaxed in their primitive cells, relative to the energy $u_0$ of two uniformly
    charged sheets in the same well, as a function of $d/r_{2D}$. The dashed blue line
    is the infinite quasicrystal energy
    where the interlayer terms self-average to zero. The quasicrystal is never the
    classical ground state.
    (b--e) Relaxed classical configurations at $d/r_{2D}=4.5$ for the honeycomb
    (red; b, d) and quasicrystal (blue; c, e) stackings in the $N=30$ square
    supercell (b, c) and the $N=78$ supercell (d, e) used in the quantum
    calculations. Filled (open) circles are electrons in the bottom (top) layer; the
    supercell is outlined with one ring of periodic images shown faintly, and
    $u-u_0$ is given under each panel. In both supercells the honeycomb remains lower,
    consistent with (a). Insets: in-plane structure factor on
    the supercell reciprocal lattice. Energies are in units of
    $\mathrm{Ha}\,(r_{2D}/\aB)^{-1}$.}
    \label{fig:sq_vs_hc_vs_qc}
\end{figure*}

In the classical limit $a_B \to 0$, the optimal bilayer stacking is determined by the competition between in-plane (intralayer) and out-of-plane (interlayer) Coulomb forces. At large interlayer separation, intralayer forces dominate, and each layer forms a triangular lattice; the layers stagger into a honeycomb pattern to minimize interlayer interaction energy. As the interlayer separation decreases, interlayer interactions drive each layer to form a square pattern, and the layers again stagger to minimize energy. In order to quantify the classical competition between the different stackings, we minimize the potential-energy terms of the Hamiltonian~\eqref{eq:H_detail} for infinite triangular and square lattices with respect to the
three-dimensional positions of the charges. In both cases, we subtract the
energy $u_0$ per electron of two uniformly charged sheets at equilibrium in the well. This extracts the stacking-dependent contribution to classical energy.
The resulting energies for the square (green) and honeycomb (red) are displayed in Fig.~\ref{fig:sq_vs_hc_vs_qc}(a) as a function of $d/r_{\rm 2D}$, and show that, as the separation increases, interlayer interactions drive each layer to form a square pattern, and the layers again stagger to minimize energy.

Finally, to further highlight the role of quantum fluctuations in the formation of the electronic quasicrystal, we compare the energy of the infinite quasicrystal with that of the honeycomb stackings. For the quasicrystal, the spatially modulated interlayer terms of
Eq.~\eqref{eq:Vl} vanish by self-averaging in the thermodynamic limit (see Sec~\ref{sec:selfa} below for more details), so that its
energy is simply that of two decoupled triangular monolayers,
independent of $d$ (dashed line). The honeycomb and quasicrystal
energies approach each other exponentially with $d$, but the quasicrystal never becomes the classical ground state.
These conclusion survive the inclusion of relaxation and finite size effect, modeled by performing gradient descent with the Hamiltonian~\eqref{eq:H_detail}, as shown in Fig.\ref{fig:sq_vs_hc_vs_qc} (b-e) for the case of the square and triangular supercells with $N=30, 70$, respectively.

\section{Self-Averaging in quasi-crystal stacking}
\label{sec:selfa}
We now comment on the self-averaging of the inter-layer Coulomb energy $E_c$ of the twisted bilayer. Consider two layer $l$ at
height $z_l$ with electrons forming a triangular lattice with lattice vectors
$\mathbf R_l$, with origin $\mathbf{s}_l$, and reciprocal lattice vectors $\mathbf G_l$. The top lattice is rotated by the twist angle $\theta$ with respect to the
bottom one, $\theta=0$ for the honeycomb stacking and $\theta=30^\circ$ for the
quasicrystal. Since each layer holds half of the electrons, the area of its unit cell
is $A_{\rm cell}=2\pi(r_{2D})^2$. By Poisson summation, the charge density of
layer $l$ and the potential it sources read
\begin{align}
    \rho_{l}(\mathbf{r}) &= \delta(z-z_l) \sum_{\mathbf{R}_{l}}
    \delta(\boldsymbol{\rho} - \mathbf{R}_l - \mathbf{s}_l)
    = \frac{\delta(z-z_l)}{A_{\rm cell}}\sum_{\mathbf{G}_{l}}
    e^{i \mathbf{G}_{l} \cdot (\boldsymbol{\rho}-\mathbf{s}_l)},\\
    V_l(\mathbf{r}) &= \frac{1}{a_B}\sum_{\mathbf{R}_l}
    \frac{1}{\sqrt{(\boldsymbol{\rho} - \mathbf{R}_{l} - \mathbf{s}_l)^2 + (z-z_l)^2}}
    = -\frac{2\pi}{a_B A_{\rm cell}} |z-z_l|
    + \frac{2\pi}{a_B A_{\rm cell}} \sum_{\mathbf{G}_{l} \neq \mathbf{0}}
    \frac{e^{-|\mathbf{G}_{l}| |z-z_l|}}{|\mathbf{G}_l|}
    e^{i \mathbf{G}_l \cdot (\boldsymbol{\rho}-\mathbf{s}_l)},
    \label{eq:Vl}
\end{align}
where the $\mathbf{G}_l = \mathbf{0}$ divergence is removed by the positively charged
background planes. In what follows we keep only the spatially modulated part and drop
the homogeneous term $-2\pi|z-z_l|/(a_B A_{\rm cell})$, which does not depend on the
stacking.

Integrating over an area $A$, the areal energy density of the charges of layer $l$ in
the potential of layer $l'$, a vertical distance $d$ away, is
\begin{align}
    E_c = \frac{1}{A} \int d^3\mathbf{r} \ \rho_{l}(\mathbf{r}) \ V_{l'}(\mathbf{r})
    = \frac{2\pi}{a_B A_{\rm cell}^2}\sum_{\mathbf{G}_{l}} \sum_{\mathbf{G}_{l'} \neq \mathbf{0}}
    \frac{e^{-|\mathbf{G}_{l'}| d}}{|\mathbf{G}_{l'}|}
    e^{-i (\mathbf{G}_{l} \cdot \mathbf{s}_{l} + \mathbf{G}_{l'} \cdot \mathbf{s}_{l'})}
    \left[ \frac{1}{A} \int_A d^2\boldsymbol{\rho} \
    e^{i (\mathbf{G}_{l} + \mathbf{G}_{l'}) \cdot \boldsymbol{\rho}} \right].
    \label{eq:selfav}
\end{align}
In the thermodynamic limit the bracket becomes $\delta_{\mathbf{G}_{l}, -\mathbf{G}_{l'}}$.
When the layers are rotated by an incommensurate angle no term survives and $E_c = 0$.
For a given pair $\mathbf{G}_{l}, \mathbf{G}_{l'}$, this self-averaging takes place
over the period $2\pi |\mathbf{G}_{l} + \mathbf{G}_{l'}|^{-1}$ of the oscillatory
integrand, which depends on $\theta$. For first-shell vectors with
$|\mathbf G_l|=|\mathbf G_{l'}|$, this period grows as the layers approach alignment as $\theta\to0$,
so that ever larger systems are needed for self-averaging. At alignment,
$\mathbf G_l=-\mathbf G_{l'}$, the integrand no longer oscillates and the inter-layer
term contributes fully, as in the honeycomb stacking: in this case, the registry shift $\mathbf{s}_{l} - \mathbf{s}'_{l}$ must be chosen so as to minimize the stacking energy. The opposite case is
$\theta=30^\circ$, where $|\mathbf G_l+\mathbf G_{l'}|$ is largest and the averaging
length is comparable to the lattice constant, so that our supercells with $N=30$ and
$78$ can already capture the incommensurate limit.

\end{document}